\documentclass[a4paper,11pt]{article}
\usepackage{jheppub}
\usepackage{amsmath,amssymb}
\usepackage{natbib}
\usepackage{graphicx,subfigure}
\usepackage{hyperref}
\usepackage{braket}
\usepackage{comment}
\usepackage[normalem]{ulem}

\newcommand{\dt}{\boldsymbol{\cdot}}
\newcommand{\Q}{\mathcal{Q}}

\newcommand{\bQ}{\boldsymbol{\mathcal{Q}}}
\newcommand{\bq}{\boldsymbol{\mathsf{Q}}}

\newcommand{\si}{\sigma}
\newcommand{\la}{\lambda}
\newcommand{\be}{\begin{equation}}
 \newcommand{\ee}{\end{equation}}
 \newcommand{\bse}{\begin{subequations}}
 \newcommand{\ese}{\end{subequations}}
\newcommand{\ba}{\begin{eqnarray}}
\newcommand{\ea}{\end{eqnarray}}

\usepackage{color}
\usepackage[normalem]{ulem}  

\begin{document}


\title{Black hole complementarity from microstate models: A study of information replication and the encoding in the black hole interior}

\author{Tanay Kibe, Sukrut Mondkar, Ayan Mukhopadhyay and Hareram Swain}
\emailAdd{tanayk@smail.iitm.ac.in, sukrut@physics.iitm.ac.in, ayan@physics.iitm.ac.in, dhareram1993@physics.iitm.ac.in }
\affiliation{Center for Operator Algebras, Geometry, Matter and Spacetime, and \\Center for Strings, Gravitation and Cosmology,\\ Indian Institute of Technology Madras, Chennai 600036, India}

\date{\today}

\abstract{
We study how the black hole complementarity principle can emerge from quantum gravitational dynamics within a local semiclassical approximation. Further developing and then simplifying a microstate model based on the fragmentation instability of a near-extremal black hole, we find that the key to the replication (but not cloning) of infalling information is the decoupling of various degrees of freedom. The infalling matter decouples from the interior retaining a residual time-dependent quantum state in the hair which encodes the initial state of the matter non-isometrically. The non-linear ringdown of the interior after energy absorption and decoupling also encodes the initial state, and transfers the information to Hawking radiation. During the Hawking evaporation process, the fragmented throats decouple from each other and the hair decouples from the throats. We find that the hair mirrors infalling information after the decoupling time which scales with the logarithm of the entropy (at the time of infall) when the average mass per fragmented throat (a proxy for the temperature) is held fixed. The decoding protocol for the mirrored information does not require knowledge of the interior, and only limited information from the Hawking radiation, as can be argued to be necessitated by the complementarity principle. We discuss the scope of the model to illuminate various aspects of information processing in a black hole.}

\maketitle
\section{Introduction}

The black hole complementarity principle \cite{PhysRevD.48.3743,PhysRevD.50.2700,Harlow:2014yka,Raju:2020smc,Kibe:2021gtw} says that if the semi-classical description of the near-horizon geometry is valid and the evaporation via Hawking radiation is an unitary process, then information of the initial state is \text{both} inside and outside of the horizon, yet these two copies cannot be simultaneously accessed by any observer. The apparently simple assumptions supporting this hypothesis have been shown to violate the strong sub-additivity of entanglement entropy \cite{Almheiri:2012rt,PhysRevLett.110.101301} (see also \cite{Mathur:2009hf}), and many arguments have been formulated in the literature for rescuing the complementarity principle from the clutches of paradoxes (see \cite{Harlow:2014yka,Raju:2020smc,Kibe:2021gtw} for some relevant reviews). In this work we explore how the complementarity principle can emerge in a way which is consistent with quantum information theory via explicit microscopic models which preserve unitarity and a local semi-classical description of the horizon. In particular, we investigate the fundamental principles behind replication of information.

Our essential object of study is how quantum matter interacts with a black hole microstate, and how such an interaction leads to replication of information in consistency with the absorption and relaxation properties of the semiclassical black hole geometry. Our results indicate that the general principle enabling the replication of information is the decoupling of the quantum matter from the black hole interior, which is inbuilt into the gravitational microstate dynamics, in the large $N$ limit where we can employ a local semi-classical description. This \textit{decoupling principle} implies that while the quantum matter typically looses energy to the black hole interior, the expectation values of the operators which couple them decay and the initial microstate of the interior relaxes to another (stationary) microstate. This relaxation process is non-linear unlike the quasi-normal mode ringdown of the classical black hole; and we find that the amplitudes, phases and decay rates of the relaxation (although the decay rate has a narrow distribution) are determined by the initial state of the matter prior to decoupling. Therefore, the initial state of the quantum matter is encoded into the von Neumann algebra describing the final ringdown phase of the microstate geometry undergoing relaxation. Remarkably, the final decoupled state of the residual quantum matter which becomes part of the hair, is time-dependent (not an energy eigenstate but with fixed expectation value of energy determined by its intrinsic Hamiltonian after decoupling) and also non-isometrically encodes the initial state. (See \cite{Akers:2022qdl} for a broad discussion on non-isometric encoding in the black hole interior.)

This implies that information is replicated twice -- one of the copies is retained by the residual time-dependent state of quantum matter and another copy is the encoding into the relaxation process of the decoupled interior that admits local semi-classical description. The second copy is in turn transferred to the Hawking quanta during Hawking evaporation. 

At the heart of our explicit local semi-classical demonstration of the emergence of the black hole microstate complementarity from the decoupling principle, is the non-linear unitary mean field channel which is justified in the large $N$ limit. Our model also has a smooth classical limit for the matter sector in which the matter reaches its ground state losing all the energy to the black hole interior losing all information, and implying that its quantum treatment is essential. Our non-linear unitary channel is reminiscent of the well studied evolution of a quantum system undergoing continuous weak measurement for which a consistent description is non-linear and state-dependent \cite{Jacobs2006}. Our study yields support for the view that the semi-classical limit of quantum gravity has inherent statistical features similar to quantum trajectories of systems undergoing continuous weak measurement. In what follows we first briefly describe our tractable microstate model and its further simplification which enables simple and explicit calculations presented in this work. Also our demonstration of emergence of the  black hole complementarity principle in the \textit{local} semi-classical approximation compliments the demonstration of the reproduction of the Page curve \cite{AEMM,PeningtonQES,AlmheiriQES,Penington:2019kki,Almheiri2020Islands} via the quantum extremal surface \cite{EngelhardtWall} for the bath collecting the Hawking radiation in the \textit{global} semi-classical approximation.

At the outset, we emphasize that our microstate model, at least in its simpler versions, is not derived and may not capture all the desired features like accounting for the Bekenstein-Hawking entropy. Nevertheless, it models the crucial information processing properties of the black hole including quantum information mirroring \cite{Hayden_2007} in a way which is consistent with the complementarity principle. Furthermore, it is possible to simulate our models by lattices of Sachdev-Yi-Kitaev (SYK) quantum dots \cite{Sachdev_1993,Kitaev_2015}. Our model, first presented in \cite{PhysRevD.102.086008}, has been inspired by the fragmentation instability of the near horizon geometry into two-dimensional anti-de Sitter ($AdS_2$) throats \cite{Brill1992,Maldacena:1998uz}. The assumption is that these throats, which are formed as a result of fragmentation, crystallize and interact with each other through the gravitational modes of the remaining original unfragmented throat which we call hair. The latter also opens towards the asymptotic part of the spacetime geometry, which for simplicity is replaced by a bath when we incorporate Hawking radiation. It was shown that such a model has a large number of stationary/static microstates. Furthermore, if one or a few throats are shocked by infalling matter, then any microstate transits to another microstate rapidly with almost all the energy absorbed by the combined Arnowitt-Deser-Misner (ADM) masses of the throats. In this paper, we show that in the thermodynamic limit, the relaxation time from one microstate to another is determined by the average mass of the fragmented throat, which is a proxy for the temperature. Therefore the model reproduces the absorption and relaxation properties of a classical black hole. 

The model demonstrates a degree of ergodicity (which is yet to be quantified appropriately), so that the final microstate is effectively obtained from the initial microstate after perturbations in a pseudo-random manner. Consequently, if one allows tunnelling of quantum matter between the throats, then the microstates can transition from one to another, and a given microstate can be expected to have an exponentially small probability that can be quantified. This feature is reminiscent of the proposed microstate dynamics in the fuzzball scenario \cite{Mathur:2005zp,Bena:2022rna}.

A key conjectured feature of a semi-classical black hole is Hayden-Preskill information mirroring \cite{Hayden_2007,Sekino:2008he}, which states that the information of the infalling qubits quickly come out of the black hole (in scrambling time). An essential feature of our microstate model is that it demonstrates Hayden-Preskill information mirroring, which was established in \cite{PhysRevD.102.086008} by approximating infalling matter as shocks inserted into individual throats. Here classical information was encoded into the sequence and location of the shocks (i.e. an ordered list). After the shocks, the initial microstate quickly relaxes to another while the hair decouples. It was found that the sequence and location of the shocks can be decoded from simple observables in the decoupled hair oscillations. Remarkably, the explicit Hayden-Preskill decoding protocol is blind to the interior (that is the same for any initial microstate) and only requires simple features of the pre-existing decoupled radiation. We will discuss later that both of these features are additional ingredients necessitated by the complementarity principle. 

Here we find that during Hawking evaporation, the fragmented throats decouple from each other and the hair decouples from the throat. The decoupled hair mirrors infalling information, with the Hayden-Preskill decoding protocol once again not requiring any knowledge of the interior and only simple measurements on the Hawking radiation. The \textit{decoupling time scales like the scrambling time}, which is the time necessary for mirroring to occur, as the logarithm of the entropy at the time of in-fall.

The rest of this paper is organized as follows. In Sec. \ref{Sec:microstate}, we review the microstate model and establish the relaxation time to be determined by the temperature. We then further develop the model and simplify it to study the replication of infalling quantum information. In Sec. \ref{Sec:Replication}, we study the replication of information in the simplified single throat model. In Sec. \ref{Sec:Hawkingrad}, we study the full model incorporating the Hawking radiation, establish grand decoupling of various degrees of freedom, and study information mirroring after decoupling time along with the explicit Hayden-Preskill decoding protocol. We also show that the decoupling time depends on the entropy like the scrambling time. Finally, in Sec. \ref{Sec:conclusion}, we conclude with a discussion about some fundamental questions which need to be addressed in the future.

\section{The mircostate model}
\label{Sec:microstate}
The microstate model proposed in \cite{PhysRevD.102.086008} is based on the fragmentation of the near-horizon geometry of near-extremal black holes into two-dimensional anti-de Sitter ($AdS_2$) throats via gravitational instantons \cite{Brill1992,Maldacena:1998uz}. This fragmentation process cannot be studied in a controlled perturbative expansion. Although each $AdS_2$ throat has a $SL(2,R)$ isometry, the latter can be explicitly broken by the interactions. Assuming that the original asymptotic region of the unfragmented throat is preserved, we can conclude that an overall diagonal $SL(2,R)$ symmetry remains unbroken.  It is expected that there will be many low energy modes which live in the asymptotic region of the original unfragmented throat as the instantons which lead to the fragmentation indeed have many zero modes \footnote{The semi-classical calculation is trust-able only when the center of the throats are sufficiently far from each other.}.  Since each $AdS_2$ throat can have an ADM mass, these throats are typically heavy degrees of freedom, while the gravitational charges of  the unbroken throat are softer and hair-like. It can be expected that the heavy throats form a dynamical lattice over which the soft hair is delocalized. Due to lack of a derivation from first principles, the simplest interacting model has been proposed in \cite{PhysRevD.102.086008} based only on the symmetries, split property of the energy in microstates in the semi-classical approximation and the phenomenological property of relaxation as described below. 

\begin{figure}
    \centering     
    \includegraphics[scale=0.6]{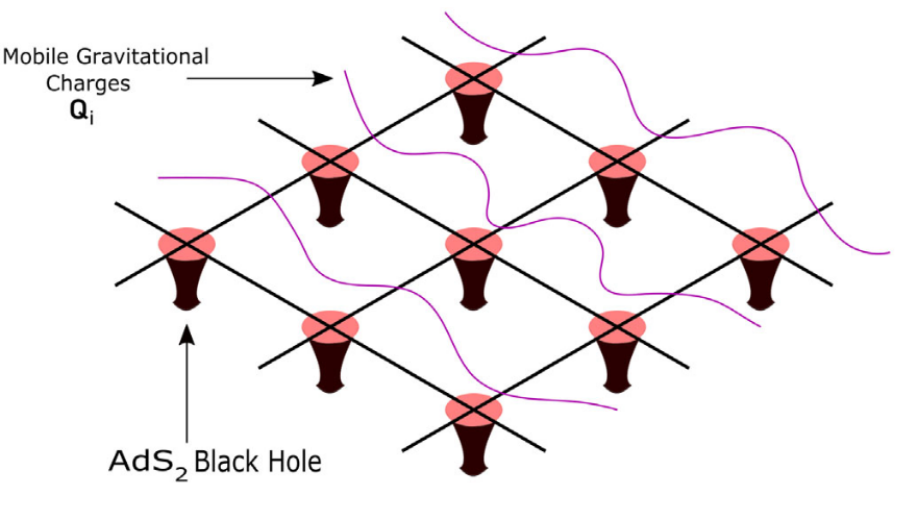}
    \caption{A schematic of the microstate model. This figure is from \cite{PhysRevD.102.086008}.}
    \label{Fig:microstate}
\end{figure}

One can also view this model as a quantum black hole simulator in which the horizon (say $S^2$ or $S^1$ for a $3+1$ or a $2+1$ dimensional black hole respectively) is discretized into a lattice, and a Sachdev-Yi-Kitaev (SYK) type quantum dot is placed at each lattice site. Holographically, in the low energy and large $N$ limit, each quantum dot can be described by Jackiw-Teitelboim (JT) gravity \cite{Teitelboim:1983ux,Jackiw:1984je,Brown:1988am} of a nearly anti-de Sitter ($NAdS_2$) throat \cite{Almheiri:2014cka,Maldacena:2016hyu,Maldacena:2016upp,Engelsoy:2016xyb,Maldacena:2019cbz,Joshi:2019wgi,Sachdev:2023try}. So, this model is an example of what can be called \textit{fragmented holography}, where different degrees of freedom (here the quantum dots) have independent dual holographic descriptions, while the interactions between them happen via other degrees of freedom, namely the hair, which are weakly self-interacting and should be treated perturbatively (without evoking any holographic description) \footnote{The latter are similar to the degrees of freedom describing a dynamical source in semi-holography.}. For an illustration, see Fig.\ref{Fig:microstate}. 

For simplicity, here we will consider a $2+1$-dimensional black hole so that the horizon is discretized $S^1$, and thus a chain with periodic boundary conditions. The main ingredients of the microstate model described on a discretized horizon are thus:
\begin{enumerate}
\item immobile $SL(2,R)$ charges $\bQ_i$ of the $AdS_2$ throats at each lattice point $i$ built out of the time-reparametrization mode $t_i(u)$ labelling the low energy states of the dual quantum dot, and 
\item propagating $SL(2,R)$ hair charges $\bq_i$ over the lattice.
\end{enumerate}

\textbf{The $SL(2,R)$ charges $\bQ_i$ and the $AdS_2$ throats:} Each $AdS_2$ throat in this model is described by Jackiw-Teitelboim (JT) gravity. The dynamical degree of freedom of each throat is a time-reparametrization mode $t_i(u)$ which essentially maps the time of a non-trivial state of the dual quantum dot to that of the vacuum. The former is denoted as $u$ and is identified with the observer's time. (The correlation functions in the non-trivial low energy state can be obtained from the vacuum correlation functions just via the time-reparametrization $t_i(u)$.) Note that $u$ is a global time-coordinate and is same for each quantum dot. The time-reparametrization at each throat is however defined only upto a local $SL(2,R)$ transformation that are the isometries of each $AdS_2$ throat which correspond to the Noether charges $\bQ_i$. More details of JT gravity will appear in Sec \ref{Sec:Setup}. For the moment, we note that the $SL(2,R)$ charges $\bQ_i$ can be constructed from the time-reparametrization mode as follows:
\begin{align}
    \bQ^+_i &=\frac{\ddot{t}_i(u)}{{\dot{t}_i(u)}^2} - \frac{{\ddot{t}_i(u)}^2}{{\dot{t}_i(u)}^3},\\\nonumber
    \bQ^0_i &= t_i(u)\left( \frac{\dddot{t}_i(u)}{{\dot{t}_i(u)}^2} - \frac{{\ddot{t}_i}^2}{{\dot{t}_i(u)}^3}\right)-\frac{\ddot{t}_i(u)}{{\dot{t}_i(u)}},\\\nonumber
    \bQ^-_i &=t_i(u)^2\left( \frac{\dddot{t}_i(u)}{{\dot{t}_i(u)}^2} - \frac{{\ddot{t}_i(u)}^2}{{\dot{t}_i(u)}^3}\right)-2 t_i(u)\left(\frac{\ddot{t}_i(u)}{{\dot{t}_i(u)}}-\frac{\dot{t}_i(u)}{{t_i(u)}}\right).
\end{align}
The ADM mass $M_i$ of the $AdS_2$ throat (which thus has a two-dimensional black hole) is simply the Casimir of these charges:
\begin{equation}
    M_i  = -\bQ_i \boldsymbol\cdot\bQ_i= -2 Sch(t_i(u),u),
\end{equation}
where $Sch$ denotes the Schwarzian derivative defined via $Sch(f(u),u) = \frac{f'''(u)}{f'(u)} - \frac{3}{2} \left(\frac{f''(u)}{f'(u)} \right)^2$and the $SL(2,R)$ invariant dot product of two $SL(2,R)$vectors $\boldsymbol{\mathcal{A}}$ and $\boldsymbol{\mathcal{B}}$ is defined via
\begin{equation}
    \boldsymbol{\mathcal{A}} \boldsymbol\cdot \boldsymbol{\mathcal{B}}:= -\boldsymbol{\mathcal{A}}^0 \boldsymbol{\mathcal{B}}^0 + \frac{1}{2}\left(\boldsymbol{\mathcal{A}}^+\boldsymbol{\mathcal{B}}^-+\boldsymbol{\mathcal{A}}^-\boldsymbol{\mathcal{B}}^+\right).
\end{equation}

\textbf{The delocalized quantum hair:} At each site we have the $SL(2,R)$ hair charges $\bq_i$ which are assumed to obey the discretized Klein-Gordon equation when they are decoupled from the localized lattice charges $\bQ_i$
\begin{equation}
    \ddot{\bq}_i - \frac{1}{\si^2}( \bq_{i-1} + \bq_{i+1} - 2 \bq_i) = 0.
\end{equation}
In what follows, we will treat $\bq_i$ classically and study quantization of the hair in future work.

\textbf{The interacting model and the microstates:} Our model is the simplest one where we introduce interactions based on these four principles:
\begin{enumerate}
    \item The individual $SL(2,R)$ symmetries of the throats should be broken to an overall diagonal $SL(2,R)$ which is the isometry of the original unfragmented throat.
    \item The total conserved energy should be the sum of the ADM masses of the throats and the kinetic energy of the hair (as in classical electromagnetism, the total conserved energy is simply the sum of the field energy and the matter energy with the interaction term $J_\mu A^\mu$ in the Lagrangian not appearing here),
    \item There should be many stationary solutions of the system, namely microstates, and the system should relax \textit{quickly} to one of these microstates after external perturbations.
    \item The energy of the hair in each microstate should further split into an interior part determined by the charges of the throats and an exterior part which should be independent of the data of the interior.
\end{enumerate}
The second property is essential for the fourth \textit{split property} to emerge. It should be noted that the fourth requirement of the split property of the energy of the hair \textit{applies to a microstate and not an arbitrary time-dependent solution of the system}. In the third requirement, the \textit{quickness} of the relaxation implies that the relaxation time should be independent of the system size when the average energy per degree of freedom is fixed. The latter is a proxy for the temperature, and therefore this requirement implies that the relaxation time should be determined by the temperature as in case of a smooth unfragmented horizon geometry. We proceed to discuss the model following \cite{PhysRevD.102.086008}, and present the new result that the relaxation time of the model is indeed determined by the average mass, and is independent of the system size, and the specific details of the microstate.

It turns out that the simplest dynamics with the four desired properties are given by these equations:
\begin{align}
 \dot{M}_i &= - \lambda\left(\bQ_{i-1}+ \bQ_{i+1}-2 \bQ_i\right)\boldsymbol{\cdot} \bq_i',\label{Eq:LatticeEoms1}\\
\ddot{\bq}_i &=\frac{1}{\sigma^2}\left( \bq_{i-1} + \bq_{i+1} - 2 \bq_i\right)+\frac{1}{ \lambda^2}\left(\bQ_{i-1}+ \bQ_{i+1}-2 \bQ_i\right)\label{Eq:LatticeEoms2}.
 \end{align}
The first equation is a type of diffusion equation for the masses of the throats, and this is responsible for the property of relaxation. The second equation is a sourced and discretized Klein-Gordon equation for the propagation of hair over the lattice. 
We can readily verify that a global $SL(2,R)$ symmetry is preserved, and also $\sum_i \dot{\bq}_i$ is a constant of motion. We do not present the action for the whole model, and the details of the gravitational dynamics in each throat since the latter can be reduced simply to \eqref{Eq:LatticeEoms1}. Briefly, the terms on the right hand side of \eqref{Eq:LatticeEoms1} produce infall of null dust in the $i$-th throat leading to the evolution of its ADM mass $M_i$ \cite{PhysRevD.102.086008}. We will give details of the gravitational dynamics in the next subsection when we simplify this model further and also introduce bulk scalar fields. We note that our model has two constants, namely $\lambda$ which is responsible for the diffusion of mass, and $\sigma$ related to the stiffness of the lattice. 

This model has a total conserved energy given by
\begin{align}
&E = E_{\bQ} + E_{\bq},\label{Eq:Enmicro1}\\
&E_{\bQ} = \sum_i M_i = M,\label{Eq:Enmicro2}\\
&E_{\bq}= \frac{\lambda^3}{2}\sum_i {\dot{\bq_i}} \boldsymbol\cdot {\dot{\bq_i}}\label{Eq:Enmicro3} + \frac{\lambda^3}{2\sigma^2}\sum_i \left(\bq_{i+1} -\bq_i\right)\boldsymbol\cdot \left(\bq_{i+1} -\bq_i\right).
\end{align}
The expressions above imply that we do satisfy our second requirement that the total energy is split into the total ADM masses of the throats and the kinetic energy of the hair including the square of the discretized spatial gradients.

Furthermore, this model indeed has a large number of stationary solutions, which are identified as the black hole microstates. One can verify that the most general stationary solutions of the model are the following up to a global $SL(2,R)$ transformation:
\begin{align}
    &\bQ_i^0 = \mathbf{Q}, \quad \bQ_i^\pm :\text{ random and constant},\label{Eq:mic1}\\
    &\bq_i^{\pm} = -\frac{\sigma^2}{\lambda^2} \bQ_i^\pm + \mathbf{K}^\pm, \quad \text{ locked to interior}\label{Eq:mic2}\\
    &\bq_i^0 = {\bq_i^0}^{mon} + {\bq_i^0}^{\rm rad}, {\bq_i^0}^{\rm rad} = q_i(u)\label{Eq:mic3} \\
    &{\bq_i^0}^{\rm mon}= \alpha  u ,\quad \text{ primordial frame}, \label{Eq:mic4}\\ 
    & \ddot{q}_i - \frac{1}{\sigma^2}(q_{i+1}+ q_{i-1}- 2 q_i) = 0, \quad \sum_i \dot{q}_i =0.\label{Eq:mic5}
\end{align}
We readily see from \eqref{Eq:mic1} that one of the components of the $SL(2,R)$ charges of the throats should be homogeneous and constant, and this component can be chosen to be the zeroth component without loss of generality by exploiting the global $SL(2,R)$ symmetry. The $\pm$ components of the charges of the throat are random constants. 

As for the hair, the $\pm$ components in the microstates are constant and locked to the corresponding constant charges of the throats up to global additive constants as shown in \eqref{Eq:mic2}. These components are thus determined fully by the configuration of the interior parameterized by the charges of the throats. The zeroth component of the hair is decoupled from the interior, and has two parts as shown in \eqref{Eq:mic3}. One part is simply the monopole term ${\bq_i^0}^{mon} $, determined by the conserved quantity $\sum_i \dot{\bq}_i$, which should be in the zeroth direction (due to the choice of the global frame) and of magnitude $\alpha$ as in \eqref{Eq:mic4}. The other part of the zeroth component of the hair is the radiation ${\bq_i^0}^{\rm rad}$, whose sum of derivatives should vanish (to remove the redundancy with the monopole term) and which satisfies free discretized Klein Gordon equation. Both the monopole and radiation terms comprising the zeroth component of the hair are thus independent of the charges of the throats, and therefore the interior. It is easy to check that \eqref{Eq:mic1}-\eqref{Eq:mic5} are solutions of the lattice equations of motion \eqref{Eq:LatticeEoms1} and \eqref{Eq:LatticeEoms2}. For a proof that these are the only stationary solutions (up to a global $SL(2,R)$ transformation) see \cite{PhysRevD.102.086008}.

There are further constraints on $\bQ_i$ from the requirement that $t_i(u)$ in each throat should be real, and their first and second derivatives should be continuous. (The latter should hold as the equations of motion \eqref{Eq:LatticeEoms1} involve the first derivative of ${M_i}$ and thus the fourth derivative of $t_i(u)$ -- the third derivatives of $t_i(u)$ are discontinuous in presence of shocks.) These imply
\begin{equation}\label{Eq:arrow1}
    Q \leq -M_i, \quad \bQ_i^\pm \leq 0, \quad \bQ_i^+ + \bQ_i^- \geq 2 Q
\end{equation}
or 
\begin{equation}\label{Eq:arrow2}
    Q \geq -M_i, \quad \bQ_i^\pm \geq 0, \quad \bQ_i^+ + \bQ_i^- \leq 2 Q.
\end{equation}
The first set of conditions lead to $\dot{t}_i\geq 0$ while the latter imply $\dot{t}_i\leq 0$. Thus a common arrow of time emerges in the microstates. We choose the former possibility as it leads to the future directed arrow of time. Finally, the set of microstate solutions are those which satisfy \eqref{Eq:mic1}-\eqref{Eq:mic5} along with the constraint \eqref{Eq:arrow1} with a total fixed energy.

It is also easy to see that in microstate solutions, the total kinetic energy of the hair $E_{\bq}$ defined in \eqref{Eq:Enmicro3} splits into two parts since the $SL(2,R)$ dot products between $\bq_i^0$ and $\bq_i^\pm$ vanish. The first part is the kinetic energy obtained from the ${\bq_i^0}^{mon}$ and ${\bq_i^0}^{rad}$ terms, which is independent of the data of the interior. The second part is the kinetic energy obtained from $\bq_i^{\pm}$ which are locked to the interior. Note here only the discretized spatial gradients contribute since $\bq_i^{\pm}$  are constants. This split property of the conserved energy into interior and exterior parts, is valid only for the microstate solutions given by \eqref{Eq:mic1}-\eqref{Eq:mic5}, and is not true when the system is undergoing relaxation. Our fourth requirement is nevertheless satisfied.

Finally, we need to check the property that the system should relax to a microstate quickly after external perturbations.  This has been studied in \cite{PhysRevD.102.086008} by shocking this system with a sequence of null shocks which are injected into the $i$-th throat at moments $u_{iA}$ and with energies $e_{iA}$ respectively. These injected energies traverse each throat along null geodesics and affect the ADM masses of the throats. As a result, the first equation of motion \eqref{Eq:LatticeEoms1} is modified to
\begin{align}\label{Eq:LatticeEoms3}
 \dot{M}_i &= - \lambda\left(\bQ_{i-1}+ \bQ_{i+1}-2 \bQ_i\right)\boldsymbol{\cdot} \dot{\bq}_i + \sum_{i,A} e_{iA} \delta(u - u_{iA}),
 \end{align}
while the second equation of motion \eqref{Eq:LatticeEoms2} remains unaltered.

If any microstate is shocked, then we find that it relaxes to another microstate given by the configuration of the $SL(2,R)$ charges of the throats along with decoupled hair radiation. Even if a single site is shocked, the final microstate is different from the initial one at sites far away from the one that is shocked, and the dynamics has characteristics of pseudorandomness which needs to be qualified more precisely. We also observe that the energy injected into the shocks is almost totally absorbed by the sum of the ADM masses of the throats while only a tiny amount (less than 1 p.c. when there are five sites) is transferred to the kinetic energy of the hair. 

\begin{figure}
    \centering     
    \includegraphics[scale=0.7]{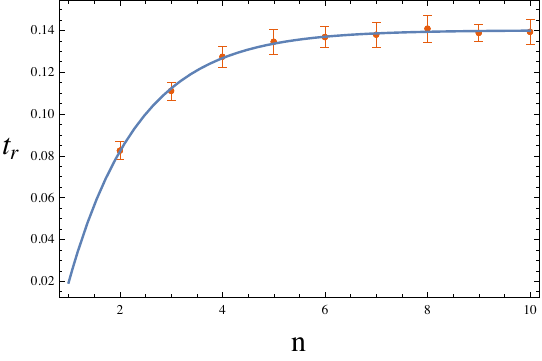}
    \caption{The relaxation time of arbitrary microstates as a function of the number of lattice sites when the average mass at each lattice site is kept fixed. The solid circles indicate the relaxation time averaged over various random initial microstates and error bars indicate the standard deviation. We note that the relaxation time is reasonably independent of the specific microstate and saturates to a constant as we increase the number of lattice sites. The data is well described by the function: $0.14-0.25 e^{- 0.73 n}$ (blue curve) with an adjusted R squared value of $0.99993$. }
    \label{Fig:relaxtime}
\end{figure}

Finally, we need to verify that the relaxation is a quick process meaning that it does not scale with the size of the system when the average mass at each lattice site is kept fixed. The relaxation time can be defined as the time when the hair decouples from the interior or equivalently when $\bQ_i^0$ homogenize to a certain desired degree of approximation implying the full system has attained its final microstate. Since the average mass gives a notion of the temperature and the relaxation time should be determined by the temperature, it is expected that the relaxation time saturates to a constant as we increase both the number of sites and also the total mass linearly with the number of lattice sites. We extract the site dependent relaxation time $t_{ri}$ by fitting $\bQ_{i-1}^0+ \bQ_{i+1}^0-2 \bQ_i^0$ to an exponentially decaying function parametrized as $ r_{i}\exp(-u/t_{ri})$. The relaxation time is defined as the mean of $t_{ri}$ as $t_r = \frac{1}{n}\sum_{i=1}^n t_{ri}$ for a lattice with $n$ sites. In Fig.\ref{Fig:relaxtime}, we note that the relaxation time of the microstates ($t_r$) in response to shocks is indeed independent of the specific microstate and depends only on the average mass at each lattice site to a good approximation.  Furthermore, the relaxation time indeed saturates to a constant as we increase the number of lattice sites keeping the average mass at each site fixed. 

To summarize, the model exhibits all the four desired properties necessary to mimic the relaxation and energy absorption properties of a classical black hole while exhibiting the split property of the energy in a microstate and the near-horizon symmetry. There are possible variants of the model in which one can connect the various throats with (Lorentzian) wormholes giving configurations which are dual to entangled chains of pure SYK spin states \cite{Kourkoulou:2017zaj} as discussed in \cite{PhysRevD.102.086008}.  It will be interesting to check the relaxation (and also scrambling) properties of these variants of this model.  In Section \ref{Sec:Hawkingrad}, we will study Hawking radiation of the microstates.

\subsection{A simplified single throat model}
\label{Sec:Setup}
The main feature of this model which leads it to reproduce the behavior of a classical black hole, especially its relaxation and energy absorption properties is that a part of the hair degrees of freedom can decouple from the throats, while another part gets locked to the charges of the throats. The throats  also decouple from each other. Essentially this leads to transition between microstates after perturbations with irreversible transfer of the injected energy to the throats. 

\begin{figure}
    \centering     
    \includegraphics[scale=0.4]{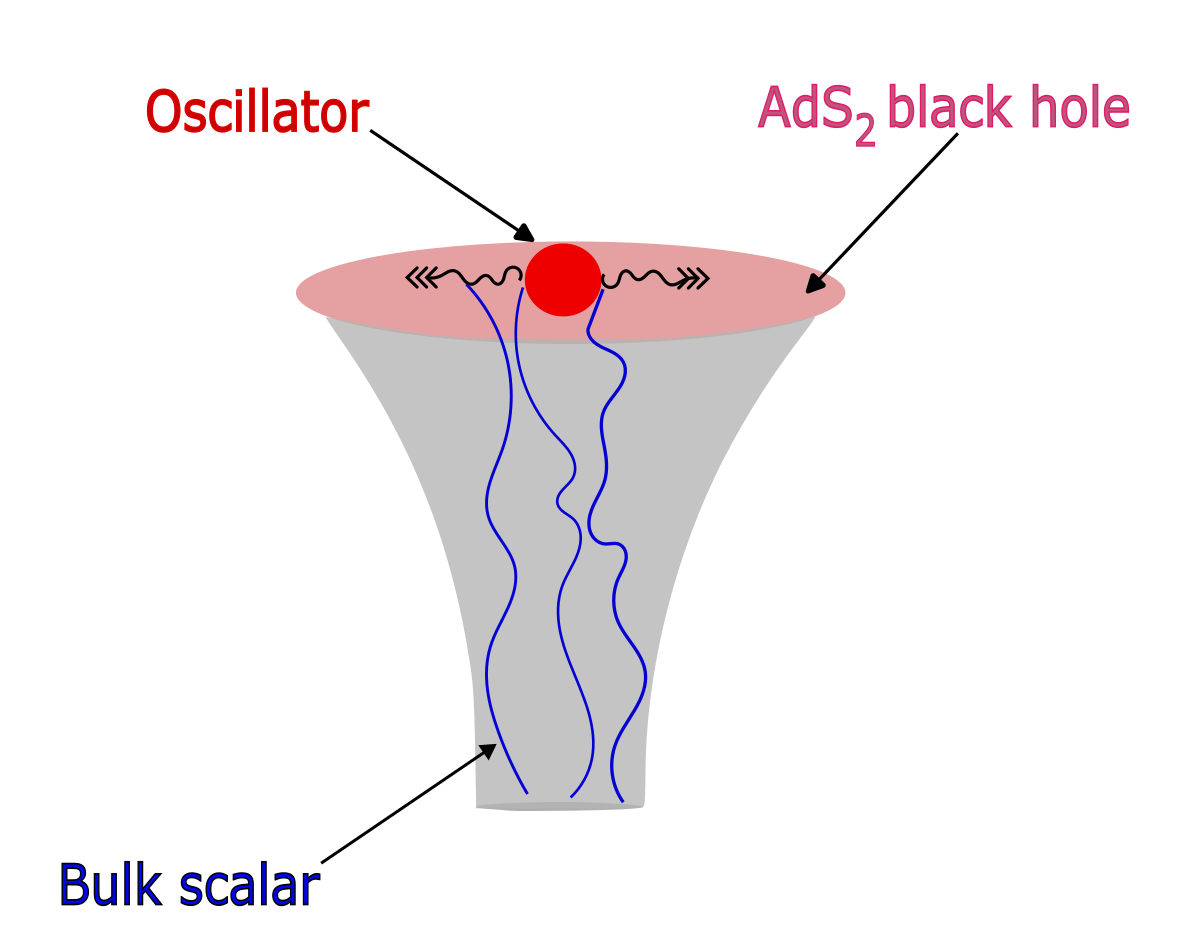}
    \caption{Schematic of the simplified single throat model}
    \label{Fig:singlethroat}
\end{figure}

This suggests that one can study a simplified setup in which (i) the incoming matter is restricted to the $s$-wave sector so that the perturbations are homogeneous, and (ii) the throats are replaced with a single one as done in dynamical mean field theory approaches (see \cite{RevModPhys.68.13} for a review). Essentially, this leads to an effective quantum harmonic oscillator (the $s$-wave sector of the infalling matter) interacting with a single throat. Anticipating that only a small residual amount of energy is left in the oscillator after the interaction with the throat, we treat the field outside the horizon (having small occupation numbers) as a quantum oscillator, but we model the wavefunction inside the throat as a classical scalar field whose backreaction on the semi-classical dilaton gravity in the throat leads to the energy transfer. A cartoon of this simplified model capturing how a quantum field interacts with a black hole microstate is shown in Fig. \ref{Fig:singlethroat}.

Concretely, let $\hat{a}^\dagger$ and $\hat{a}$ be the creation and annihilation operators of the $s$ wave mode of frequency $\omega$ respectively, and
\begin{equation}\label{Eq:x-p-def}
    \hat{x} = \frac{1}{\sqrt{2\omega}} \left( \hat{a}^\dagger+\hat{a} \right),  \quad \hat{p} = i\frac{\sqrt{\omega}}{\sqrt{2}} \left( \hat{a}^\dagger-\hat{a}\right).
\end{equation}
Therefore $\hat{x}$ is essentially the $s$-wave field mode living at the boundary of the near-horizon region and $\hat{p}$ is its conjugate momentum. Also $$\omega \left(\hat{a}^\dagger \hat{a} + \frac{1}{2}\right) = \frac{\hat{p}^2}{2} + \omega^2 \frac{\hat{x}^2}{2}.$$The action giving the Schr\"{o}dinger equation for the quantum harmonic oscillator wave function $\ket{\psi}$ is 
\begin{equation}\label{Eq:SQHO}
    S_{QHO} = \int {\rm d}u \,\, \bra{\psi}\left(i\partial_u - \omega \left(\hat{a}^\dagger \hat{a} + \frac{1}{2}\right) -  J_{HO} \hat{x} \right){\ket{\psi}}.
\end{equation}
We note that the oscillator has a source $J_{HO}$ which is needed to couple it to the bulk scalar field. Here $J_{HO}$ is an auxiliary field which will be determined in terms of the bulk scalar field as shown below.

The on-shell gravitational action of the single throat gives the partition function of the dual quantum dot theory. The gravitational fields are the metric $g$ and the dilaton $\Phi$ which together capture the time-reparametrization mode  of the quantum dot. Additionally there is a scalar field $\chi$ which represents the infalling $s$-wave inside the throat and which is dual to an operator $O$ in the quantum dot theory. The mass $m$ of $\chi$ can be obtained from the scaling dimension $\Delta$ of $O$ via $m^2l^2 = \Delta(\Delta -1)$, where $l$ is the radius of $AdS_2$. The logarithm of the partition function $Z_{QD}$ of the dual quantum dot theory is identified with the on-shell gravitational action via the holographic dictionary:
\begin{eqnarray}\label{Eq:Hol-Dic1}
     W_{QD}[J_O] &=& \ln Z_{QD}[J_O] \nonumber\\ &=& S_{\rm grav}^{\rm on-shell}[\chi^{(0)}= C_\Delta J_O].
\end{eqnarray}
Above $\chi^{(0)}$ is the leading term (a.k.a. non-normalizable mode) of the bulk scalar field $\chi$, i.e.
\begin{equation}\label{Eq:Hol-Dic2}
    \chi^{(0)} = \lim_{r\rightarrow0} r^{\Delta -1} \chi 
\end{equation}
where $r$ is the radial coordinate of the bulk spacetime whose asymptotic boundary is at $r=0$. By the holographic dictionary, $\chi^{(0)}$ is identified with the source $J_O$ upto a constant $C_\Delta$ which for reasons to be clear later we set as
\begin{equation}\label{Eq:Hol-Dic3}
    C_\Delta = - 2 (2\Delta -1), \,\, {\rm so}\,{\rm that} \,\, \chi^{(0)}= - 2 (2\Delta -1) J_O.
\end{equation}
For reasons to be explained later, we would impose $1/2< \Delta < 3/2$.

The action for the full system is 
\begin{equation}\label{Eq:FullAc}
    S = S_{QHO}[J_{HO}] + W_{QD}[J_O] - \frac{1}{\gamma}\int{\rm d}u \, J_O J_{HO}.
\end{equation}
where $\gamma$ a coupling constant, $S_{QHO}$ is given by \eqref{Eq:SQHO} and $W_{QD}$ is defined in \eqref{Eq:Hol-Dic1}. Since the mass dimensions of $J_{HO}$ and  $J_O$ are  $3/2$ and $1 -\Delta$ respectively, it follows that the mass dimension of $\gamma$ is $3/2 - \Delta$. Varying the full action w.r.t. $J_{QHO}$ and $J_{O}$ we obtain
\begin{align}\label{Eq:Sources}
    & J_O =\gamma \frac{\delta S_{QHO}}{\delta J_{HO}} = \gamma \bra{\psi} \hat{x}\ket{\psi},\nonumber\\
    &J_{HO} =\gamma \frac{\delta W_{QD}}{\delta J_{O}} = \gamma O.
\end{align}
See \cite{Ecker:2018ucc,Mondkar:2021qsf} for similar actions for semi-holography.

The gravitational action defining $W_{QD}$ is two-dimensional JT gravity \cite{Teitelboim:1983ux,Jackiw:1984je,Brown:1988am} with a negative cosmological constant minimally coupled to the bulk matter field $\chi$ as follows:
\begin{align}\label{Eq:Grav-Action}
    &S = \frac{1}{16 \pi G} \int d^2x \sqrt{-g} \left(R + \frac{2}{l^2}\right) \Phi  + S_{\text{matter}}[g,\chi] + \frac{1}{8 \pi G} \int du \sqrt{-h}\phi_b K + S_{ct}[\chi],\nonumber\\
    &S_{\rm matter} = \int{\rm d}^2 x\, \left(-\frac{1}{2}g^{\mu\nu}\partial_\mu\chi\partial_\nu \chi - \frac{1}{2}m^2 \chi^2\right),
\end{align}
Above, $l$ is the radius of $AdS_2$. Note that the dilaton $\Phi$ does not couple directly to the bulk scalar field $\chi$ and $\Phi_b$ is the boundary value of the dilaton. Recall that $u$ is the boundary time. $S_{c.t.}$ is the counterterm action  \cite{Skenderis1} at the boundary needed for obtaining a finite on-shell gravitational action where the counterterms are consistent with the Dirichlet boundary condition for $\chi$ (the latter is given by our identification of $\chi^{(0)}$ with $J_O$ via \eqref{Eq:Hol-Dic3}).

Varying the gravitational action we obtain the following equations of motion
\begin{align}
        &R +\frac{2}{l^2} = 0, \label{Eq:EOMJT1}\\
        & g_{\mu \nu} \left(\nabla^2 -\frac{1}{l^2}\right)\Phi-\nabla_\mu \nabla_\nu \Phi 
        \nonumber\\
        &=16\pi G\left(\partial_\mu\chi \partial_\nu \chi - \frac{1}{2} g_{\mu\nu} g^{\alpha\beta}\partial_\alpha\chi\partial_\beta \chi - \frac{1}{2} g_{\mu\nu} m^2 \chi^2\right),\label{Eq:EOMJT2}\\
        &(\nabla^2 - m^2)\chi = 0 
    \label{Eq:KG}
\end{align}
The dilaton acting like a Lagrange multiplier in the gravitational action \eqref{Eq:Grav-Action} imposes that the metric is always locally $AdS_2$, i.e. \eqref{Eq:EOMJT1}. The second equation \eqref{Eq:EOMJT2} above should be understood as a constraint equation. The left hand side of this equation has zero divergence in the locally $AdS_2$ metric (like in the case of the Bianchi identity for the usual Einstein's equations), and so does the right hand side when the bulk scalar field satisfies the Klein-Gordon equation \eqref{Eq:KG} as it is the energy-momentum tensor of this scalar field up to an overall constant. In order to fully determine the holographic gravitational sector, we set the boundary condition for the dilaton via 
\begin{equation}\label{Eq:BCPhi}
    \lim_{r\rightarrow0}r\Phi = \Phi_r,
\end{equation}
with $\Phi_r$ a constant. A better way to do this is to parametrize the boundary in terms of bulk coordinates $r$ and $u$ as $r = \epsilon \dot{t}(u) + \mathcal{O}(\epsilon^2)$ so that the induced metric on the boundary is fixed to $h_{tt} = - 1/\epsilon^2$. The latter amounts to the Dirichlet boundary condition for the bulk metric which implies that the physical metric for the state in the dual theory is given by ${\rm d}s^2 = - {\rm d}t^2 = - \dot{t}(u)^2 {\rm d}u^2$, so that $t(u)$ is indeed the time-reparametrization of the vacuum time which leads to the observables of this non-trivial state. Recalling that $\Phi_b(u)$ is the boundary value of $\Phi$, we note that \eqref{Eq:BCPhi} implies that $\Phi_b(u) \sim \Phi_r/\epsilon$. A finite on-shell gravitational action is obtained in the limit $\epsilon \rightarrow 0$ (see \cite{Joshi:2019wgi} for a discussion). In the dual quantum dot theory,
\begin{equation}\label{Eq:N}
    \frac{\Phi_r}{16\pi G} \sim N^2,
\end{equation}
with $N^2$  parameterizing the number of elementary degrees of freedom. 

It is useful to understand the time-reparametrization better from the perspective of an improper bulk diffeomorphism which affects the boundary. In the ingoing Eddington-Finkelstein gauge, the general form of the locally $AdS_2$ metric (satisfying \eqref{Eq:EOMJT1}) is
\begin{equation}
     {\rm d}s^2 = -2 \frac{l^2}{r^2} {\rm d}u {\rm d}r - \left(\frac{l^2}{r^2} - M(u) l^2 \right) {\rm d}u^2,
     \label{Eq:metricEF}
\end{equation}
where $M(u)$ is a time-dependent black hole mass. This metric can be transformed to pure $AdS_2$ with no black hole, i.e. to the metric
\begin{equation}
    {\rm d}s^2 = -2 \frac{l^2}{\rho^2} {\rm d}t {\rm d}\rho - \frac{l^2}{\rho^2} {\rm d}t^2,
\end{equation}
via the diffeomorphism
\begin{equation}\label{Eq:Diffeo1}
    t= t(u), \quad \rho = \frac{\dot{t}(u) r}{1-\frac{\ddot{t}(u)}{\dot{t}(u)} r},
\end{equation}
where
\begin{equation}\label{Eq:M-Sch}
    M(u) = -2 Sch(t(u),u),
\end{equation}
with $Sch$ denoting the Schwarzian derivative. It is clear then that $t(u)$ is the time-reparametrization at the boundary and $M(u)$ is the cost of the related residual bulk diffeomeorphism which preserves the Eddington-Finkelstein gauge (one can of course define it similarly in any other gauge). Note that $t(u)$ is defined up to a $SL(2,R)$ transformation which leaves the Schwarzian derivative invariant. These $SL(2,R)$ transformations therefore lead to the local isometries of the $AdS_2$ metric. Physically inequivalent states can be parametrized by the $SL(2,R)$ invariant $M(u)$, or equivalently the coset ${\rm Diff}/SL(2,R)$. 

Explicitly, the Klein-Gordon equation for the bulk scalar field (with $m^2 l^2=\Delta (\Delta-1)$) in the bulk metric \eqref{Eq:metricEF} is
\begin{equation}
    \partial_r\left(\partial_u \chi -\frac{1}{2}\left(1- r^2 M(u)\right) \partial_r \chi \right) + \frac{\Delta (\Delta-1)}{2 r^2} \chi =0.
    \label{Eq:KGscalar}
\end{equation}
which leads to the boundary expansion
\begin{equation}\label{Eq:ScalarExp}
    \chi(r,u) \approx \chi^{(0)}(u) r^{1-\Delta} +\dot{\chi}^{(0)}(u) r^{2-\Delta} + O(u)r^{\Delta} + \hdots,
\end{equation}
where $O$ should be identified with the expectation value of the dual operator in the state dual to the gravitational solution via the holographic dictionary (this follows from  $\frac{\delta W_{QD}}{\delta J_{O}}=O$, and \eqref{Eq:Hol-Dic1} and \eqref{Eq:Hol-Dic2} -- for specific $\Delta$, there can be additional local time-derivatives of $\chi^{(0)}$ contributing to $O$). 

The explicit solution to the equation of motion for the dilaton \eqref{Eq:EOMJT2} is for $\Delta = \frac54$ \cite{Joshi:2019wgi}
\begin{align}\label{Eq:dilaton_soln}
    &\Phi = \frac{\Phi_r}{r} +16\pi G\Bigg( \frac{\chi^{(0)}(u)^2}{4} r^{-1/2} + \int_0^r dr'' \frac{1}{{r''}^2} \left( \frac{1}{8}\chi^{(0)}(u)^2 {r''}^{\frac{1}{2}} - \int_0^{r''} dr' {r'}^2 \left(\partial_{r'} \chi(r',u)\right)^2\right)\Bigg).
\end{align}
with the further requirement that the mass function $M(u)$ should satisfy
\begin{equation}\label{Eq:WI1}
    \dot{M}(u) = \frac{16\pi G}{\Phi_r}\left(\Delta\, O(u) \dot{J_O}(u) + (\Delta-1) \dot{O}(u) J_O(u) \right).
\end{equation}
Above we have used \eqref{Eq:Hol-Dic3} to substitute $\chi^{(0)}$ with $J_O$.  Defining the ADM mass of the bulk spacetime to be
\begin{equation}\label{Eq:MADM}
    M_{\rm ADM} = \frac{\Phi_r}{16\pi G} M(u) - (\Delta-1) O(u) J_O(u),
\end{equation}
we can rewrite \eqref{Eq:WI1} in the form
\begin{equation}\label{Eq:EOMmass}
    \dot{M}_{\rm ADM}(u) = O(u)\dot{J}_O(u)
\end{equation}
implying that $M_{ADM}$ can be identified with the Hamiltonian of the dual quantum dot. The on-shell gravitational action in the limit $\epsilon\rightarrow 0$ turns out to be \cite{Joshi:2019wgi}
\begin{align}\label{Eq:Sgrav}
   &S_{\text{grav}}^{\text{on-shell}} = \int {\rm d}u\, M_{\rm ADM}(u)\nonumber\\
   &= \frac{\Phi_r}{16\pi G}  \int {\rm d}u\,M(u) - (\Delta-1) \int {\rm d}u\, O(u) J_O(u).
\end{align}
We note that \eqref{Eq:WI1} (or equivalently \eqref{Eq:EOMmass}) can be derived from this action by varying it w.r.t. $t(u)$ \footnote{To see this we need to use \eqref{Eq:M-Sch}, and also $O(u) = t'(u)^\Delta O_p(t(u))$ and $J_O(u) = t'(u)^{1-\Delta} J_{pO}(t(u))$. Note $J_{pO}$ and $O_p$ are sources and expectation values respectively in pure $AdS_2$ geometry with vanishing mass.}.

Furthermore, the explicit solution  \eqref{Eq:dilaton_soln} implies that for the dilaton not to diverge near the boundary faster than $r^{-1}$, we would require that $\Delta < 3/2$. However, we would need $\Delta> 1/2$ so that we are above the Breitenloher-Freedman bound and also for \eqref{Eq:Hol-Dic3} to make sense. Therefore, we restrict to $1/2< \Delta < 3/2$. 

With the above discussion on the gravitational sector, we can derive the equations of motion of the full system from the complete action \eqref{Eq:FullAc}. It is useful to first vary with respect to the auxiliary sources which lead to \eqref{Eq:Sources} determining their values as mentioned above. Then varying with respect to the wavefunction of the oscillator we obtain the Schr\"{o}dinger equation
\begin{align}\label{Eq:Schro}
\left(i\partial_u  - \omega \left(\hat{a}^\dagger \hat{a} + \frac{1}{2}\right) - \gamma O \hat{x}\right)\ket{\Psi}  
=0.
\end{align}
It follows from the discussion above that the gravitational sector can be fully determined by (i) the equation \eqref{Eq:WI1} which can be rewritten as
\begin{equation}\label{Eq:WI2}
    \dot{M}(u) = \frac{16\pi G}{\Phi_r}\gamma\left(\Delta O(u) \langle \dot{\hat{x}}(u)\rangle+ (\Delta-1) \dot{O}(u) \langle {\hat{x}}(u)\rangle \right),
\end{equation}
where $\langle {\hat{x}}(u)\rangle \equiv \bra{\Psi}\hat{x}\ket{\Psi}$ and we have used \eqref{Eq:Sources} and (ii) the Klein-Gordon equation \eqref{Eq:KGscalar} from where $O(u)$ should be extracted via the asymptotic expansion \eqref{Eq:ScalarExp} after imposing the boundary condition \eqref{Eq:Hol-Dic3}. Thus \eqref{Eq:Schro}, \eqref{Eq:WI2} and \eqref{Eq:KGscalar} together with the boundary condition \eqref{Eq:Hol-Dic3} determine the full system once the initial conditions are specified.

In order for the semiclassical gravitational description to be valid, we require the large $N$ limit where $N$ is given by \eqref{Eq:N}. For the backreaction of the quantum oscillator to be significant, we need
\begin{equation}
    \gamma = \mathcal{O}(N), \,\, \omega = \mathcal{O}(N^2).
\end{equation}
It follows from the boundary condition \eqref{Eq:Hol-Dic3} and the asymptotic expansion \eqref{Eq:ScalarExp} that the first condition above implies that $$J_O,O = \mathcal{O}(N),$$and the second condition above then implies that the energy of the oscillator is $\mathcal{O}(N^2)$.  One can interpret this as follows. Although our model describes the near horizon region, the energy of the oscillator is measured by an observer far away from the horizon just like $M_{\rm ADM}$. The redshift factor near the horizon implies that for the oscillator to have finite energy at the horizon, it should have large energy for the asymptotic observer. However, instead of being infinite as in the classical black hole geometry, the latter is $\mathcal{O}(N^2)$ and proportional to the total mass of the microstate. It would be interesting to derive this feature of our single throat model from a fuzzball perspective along with a better way to represent the asymptotic region connected to the near horizon region.  

It is easy to see from the full equations of motion that the conserved energy of the full system $E_{tot}$ is simply the sum of the energy of the oscillator
\begin{equation}\label{Eq:Eosc}
    E_{osc} = \omega \left(\langle \hat{a}^\dagger \hat{a}\rangle + \frac{1}{2}\right)
\end{equation}
and the ADM mass of the gravitational system given by \eqref{Eq:MADM}, i.e.
\begin{equation}
    E_{tot} = E_{osc} + M_{\rm ADM}.
\end{equation}
To see this we note that it follows from the Schr\"{o}dinger equation \eqref{Eq:Schro} that 
\begin{equation}
    \dot{E}_{osc}= - \gamma O \langle \dot{\hat{x}}\rangle = - O \dot{J}_O.
\end{equation}
Together with \eqref{Eq:EOMmass} the above implies that $\dot{E}_{tot} = 0$.

Since $M_{\rm ADM}$ has the interpretation of being the expectation value of the Hamiltonian of the dual quantum dot system (that represents the interior of the black hole) as discussed above, it follows that the total energy has the desired split property of being just the sum of the gravitational ADM mass and the expectation value of the Hamiltonian of the boundary system. The interaction term $\gamma\langle {\hat{x}}\rangle O$ does not appear as in the case of the full microstate model discussed before. Nevertheless, for the sake of convenience of our later discussions, we will define
\begin{equation}\label{Eq:Eint-def}
    E_{int} =- (\Delta -1) O J_O = -\gamma (\Delta -1) O \langle {\hat{x}}\rangle,
\end{equation}
and the black hole mass $M_{\rm BH}$ via
\begin{equation}
    M_{\rm BH} = \frac{\Phi_r}{16\pi G} M,
\end{equation}
so that we can split the ADM mass and the total conserved energy of the system as
\begin{equation}
    M_{\rm ADM} = M_{\rm BH} + E_{int}, \,\, E_{tot} = E_{osc} + M_{\rm BH} + E_{int}.
\end{equation}

\section{Quantum information replication in the single throat model}\label{Sec:Replication}
The primary aim of this paper is to understand what happens to the quantum information of infalling matter, and therefore we will study the single throat model introduced in the previous section in the semiclassical approximation. Intriguingly, we will find replication of the information with one copy encoded into the late time (non-linear) ringdown of the black hole mass and another copy encoded into the final quantum trajectory of the residual modes with support outside of the horizon and which decouple from the interior. Both encodings will be non-isometric. 

In order to solve the single throat model, we will need to solve the Schr\"{o}dinger equation \eqref{Eq:Schro} for the oscillator giving the s-mode of the infalling matter just outside the horizon, the equation \eqref{Eq:WI2} for the mass of the black hole, and the Klein-Gordon equation \eqref{Eq:KGscalar} for the classical bulk scalar field to extract the expectation value of the operator $O$ that couples to the oscillator using \eqref{Eq:Hol-Dic3} and \eqref{Eq:ScalarExp}. All these three equations should be solved simultaneously. In this section we will not set $\hbar = 1$ in order to compare our results with the classical limit for the oscillator.

Let us begin with the Schr\"{o}dinger equation \eqref{Eq:Schro} for which the explicitly time-dependent Hamiltonian is (see \eqref{Eq:x-p-def})
\begin{equation}\label{Eq:Hosc}
	\hat{H}(u) = \frac{1}{2} \hat{p}^{2} +\frac{1}{2} \omega^2 \hat{x}^2  + \gamma O(u) \hat{x}.
\end{equation}
This is simply a damped harmonic oscillator with a time dependent damping factor. 

To study the time evolution we can use the so-called method of invariants where we need to find a dynamically invariant operator $\hat{I}$ \cite{PhysRevLett.18.510,choi2004coherent} such that it satisfies
\begin{equation}\label{Eq:I-def}
	\frac{d \hat{I}}{du} \equiv \frac{\partial \hat{I}}{\partial u} +\frac{i}{\hbar} [\hat{H}, \hat{I}]=0.
\end{equation}
Note that the above equation is the same as the evolution of an operator in the Heisenberg picture, but here we are in the Schr\"{o}dinger picture. Let us assume that $\ket{\psi_n}$ is a (time-dependent) eigenvector of the invariant $\hat{I}$ with eigenvalue $c_n$, i.e. \begin{equation}\label{Eq:E-I}
    \hat{I}\ket{\psi_n} = c_n\ket{\psi_n}.
\end{equation}
Taking the time-derivative of both sides of this eigenvalue equation and using \eqref{Eq:I-def} we obtain 
\begin{equation*}
   -\frac{i}{\hbar} \hat{H}\hat{I}\ket{\psi_n} + \frac{i}{\hbar}\hat{I} \hat{H}\ket{\psi_n} + \hat{I} \dot{\ket{\psi_n}} = c_n \dot{\ket{\psi_n}} + \dot{c_n}\ket{\psi_n}.
\end{equation*} Using the eigenvalue equation \eqref{Eq:E-I} and since the time-dependence of $\ket{\psi_n}$ should follow from the Schr\"{o}dinger equation with the Hamiltonian $\hat{H}$ above, we obtain that
\begin{equation*}
    0 = \dot{c_n}.
\end{equation*}So we find that the eigenvalues of $\hat{I}$ are time-independent (although $\hat{I}$ could be explicitly time-dependent) and the corresponding eigenvectors with appropriate time-dependent phases give a complete basis with each element of the basis being a solution of the Schr\"{o}dinger equation with the Hamiltonian $\hat{H}$ which can be explicitly time-dependent.

Substituting the explicitly time-dependent Hamiltonian \eqref{Eq:Hosc} into the equation \eqref{Eq:I-def}, we obtain the required dynamical invariant
\begin{equation}
	\hat{I} = \frac{1}{2}  \left( (\hat{p} - p_p(u))^2 + \omega^2 (\hat{x} - x_p(u))^2\right),
\end{equation}
where $x_p(u)$ and $p_p(u)$ are functions (not operators) which satisfy the equations of motion of the classical oscillator, namely
\begin{align}
	\ddot{x_p} + \omega^2 x_p = -\gamma O(u), \quad
	{p_p} = \dot{x_p}.
 \label{Eq:class_xp_pp}
\end{align}
We can define new raising and lowering operators
\begin{align}
	\hat{b} &= \frac{1}{\sqrt{ 2\hbar}} \left( (\hat{x} - x_p(u)) \sqrt{\omega} +i \frac{(\hat{p} -p_p(u))}{\sqrt{ \omega}}  \right),\\
	\hat{b}^\dag &= \frac{1}{\sqrt{ 2 \hbar}} \left( (\hat{x} - x_p(u)) \sqrt{\omega} - i \frac{(\hat{p} -p_p(u))}{\sqrt{ \omega}}  \right),
\end{align}
so that
\begin{equation}
	\hat{I} = \hbar\omega\left( \hat{b}^\dag \hat{b} +\frac{1}{2} \right).
\end{equation}
We note that $\hat{I}$ indeed has \textit{time-independent} eigenvalues, namely $$\hbar \omega \left(n + \frac{1}{2}\right).$$

The eigenvectors of the invariant operator $\hat{I}$ are
\begin{align}\label{Eq:phin-def}
	\braket{x | \phi_{n}(u;x_p(u), p_p(u))}  &= \left(\frac{\omega}{\pi}\right)^{1/4} \frac{1}{\sqrt{2^n n!}} H_n\left[\sqrt{\frac{\omega}{\hbar}} (x-x_p(u))\right] \\
 & \quad \times \exp\left( i \frac{p_p(u)}{\hbar} x -\frac{\omega}{2\hbar} (x -x_p (u))^2\right) \exp(-i F_n(u)),
\end{align}
where $H_n$ is the $n^{th}$ order Hermite polynomial, and
\begin{equation}\label{Eq:Fn-def}
	 F_n(u) = \hbar\omega u \left(n + \frac{1}{2} \right) + \hbar\int_{0}^{u} du' \left(\frac{1}{2} \dot{x}_p(u')^2 -\frac{1}{2} \omega^2 x_{p}(u')^2 \right).
\end{equation}
Note the last line above corresponds to $\hbar$ times the classical action for the phase space trajectory given by $x_p(u)$ and $p_p(u)$. The choice of phases given by $F_n(u)$ ensure that each $\ket{\phi_n}$ is a solution of the time-dependent Schr\"{o}dinger equation \eqref{Eq:Schro}.  Note that these are simply the usual quantum oscillator eigenstates when $x_p(u) =0$ and $p_p(u) =0$; however the latter is possible only in absence of driving. One can readily check that $\braket{\phi_{n}(u) | \hat{x} | \phi_{n}(u)} = x_p(u)$ and $\braket{\phi_{n}(u) | \hat{p} | \phi_{n}(u)} = p_p(u)$. For $n=0$, we obtain the coherent state. The states $\ket{\phi_n(u)}$ for $n>0$ can be called excited coherent states. Both in the presence and absence of driving, the states $\ket{\phi_n(u)}$ provide a complete basis of states with each element of this basis solving the time-dependent Schr\"{o}dinger equation, and with the expectation values of the position and momentum, given by $x_p(u)$ and $p_p(u)$ respectively, corresponding to a specific classical trajectory in phase space since they solve \eqref{Eq:class_xp_pp}. We emphasize that the phase space trajectory $x_p(u)$ and $p_p(u)$ is the same for all $\ket{\phi_n(u)}$.

We choose initial conditions
\begin{equation}\label{Eq:x-p-in}
    x_p(0) = 0, \quad p_p(0) = 0
\end{equation}
so that the states $\ket{\phi_n}$ coincide with the usual Fock states of a free oscillator at $u=0$ as this allows us to specify the initial state of the oscillator in the basis of the usual Fock states. 

For a concrete example, let us set the initial state of the oscillator at $u=0$ to be within the qubit subspace spanned by the first two Fock space states $\ket{0}$ and $\ket{1}$ (the ground state and first excited state of the free oscillator) so that 
\begin{equation}\label{Eq:psi-in}
	\ket{\psi(u=0)} = \cos\theta e^{i\phi} \ket{0}+ \sin\theta \ket{1}.
\end{equation}
The angles $\theta$ and $\phi$ parametrize the Bloch sphere. It then follows that the time dependent state of the quantum oscillator interacting with the black hole is simply
\begin{equation}\label{Eq:psi-sol}
	\ket{\psi(u)} = \cos\theta e^{i\phi} \ket{\phi_0(u;x_p(u),p_p(u))}+ \sin\theta  \ket{\phi_1(u;x_p(u),p_p(u))}.
\end{equation}
The above solves the Schr\"{o}dinger equation \eqref{Eq:Schro} with the initial condition \eqref{Eq:psi-in} given that $x_p(u)$ and $p_p(u)$ solve the classical oscillator equations \eqref{Eq:class_xp_pp} with initial conditions \eqref{Eq:x-p-in}.

In order to see how the oscillator transfers energy to the black hole we need $\braket{x(u)}=\braket{\psi (u)| \hat{x}| \psi(u)}$ which has to be substituted into the equation for the evolution of the black hole mass given by \eqref{Eq:WI2}. Using the explicit time-dependent state \eqref{Eq:psi-sol}, we readily obtain
\begin{equation}\label{Eq:ex-xu}
    \braket{x(u)}=\braket{\psi (u)| \hat{x}| \psi(u)} = x_p(u)+\sqrt{\frac{\hbar}{2  \omega}}\sin(2\theta)\, \cos(\phi + \omega u).
\end{equation}
    
Finally, we can solve the full system by just solving the classical equations \eqref{Eq:class_xp_pp} for $x_p$ and $p_p$ with initial conditions \eqref{Eq:x-p-in} together with the equation for $M(u)$, the black hole mass, given by \eqref{Eq:WI2}, and and the Klein-Gordon equation for the bulk scalar field \eqref{Eq:KGscalar} from which we need to extract the operator $O$ which couples to $\braket{x(u)}$ via \eqref{Eq:ScalarExp}. We note that we need to use \eqref{Eq:ex-xu} both in \eqref{Eq:WI2} and also to set the boundary condition for the bulk scalar field via \eqref{Eq:Hol-Dic3}. The numerical algorithm used to implement this procedure is summarized in Appendix.\ref{Sec:AppendixB}.

The general results are as follows.
\begin{itemize}
    \item Energy is transferred from the quantum oscillator to the black hole but the oscillator retains some residual energy. This residual energy vanishes only in the classical limit $\hbar\rightarrow0$ in which complete transfer of energy from the oscillator to the black hole can be achieved.
    \item The oscillator and the black hole \textit{decouple} from each other eventually so that $\braket{x(u)}\rightarrow0$, $\braket{p(u)}\rightarrow0$ and $O(u)\rightarrow0$ at late time.
    \item Although $\braket{x(u)}\rightarrow0$ and $\braket{p(u)}\rightarrow0$, the quantum oscillator does not go to the ground state mimicking the classical limit where it goes to equilibrium. At late time, the decoupled quantum oscillator follows a quantum trajectory which \textit{encodes its initial state completely but non-isometrically}.
    \item The information of the initial state of the oscillator is also encoded into the late time ringdown of the black hole mass which settles down to a constant value eventually. Therefore, this is a transitory copy of information which can be encoded into the Hawking radiation and hair degrees of freedom.
 \end{itemize}

\begin{figure}
        \centering
        \includegraphics[scale=0.9]{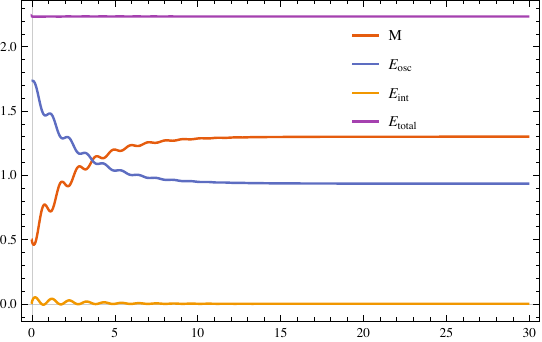}
        \caption{$E_{\text{osc}} =  \frac{1}{2} \braket{p^2} + \frac{1}{2} \omega^2 \braket{x^2}$, $E_{\text{int}} =  -(\Delta - 1) \gamma O \braket{x} $ and $M(u)$ as functions of time. Note that the total energy $E_{tot}$ is conserved. We have chosen $\omega =3.2 N^2$, $\gamma = -0.3 N$ (with $N$ given by \eqref{Eq:N}) and $\Delta=\frac{5}{4}$. Here all energies have been scaled with $N^{-2}$.}
        \label{Fig:energy_quant_osc}
\end{figure}

A typical evolution of energies for this system is shown in Fig.\ref{Fig:energy_quant_osc}.
The accuracy of the numerical algorithm is reflected in the conservation of the total energy to the desired level of precision. We indeed find that some but not all of the energy from the oscillator $E_{osc}$ (defined in \eqref{Eq:Eosc}) is absorbed by the black hole. The black hole mass and the energy of the oscillator settle down to constant values, while the interaction energy \eqref{Eq:Eint-def} decays to zero indicating eventual decoupling of the oscillator from the black hole.

\begin{figure}
\centering     
\subfigure[]{\includegraphics[scale=0.7]{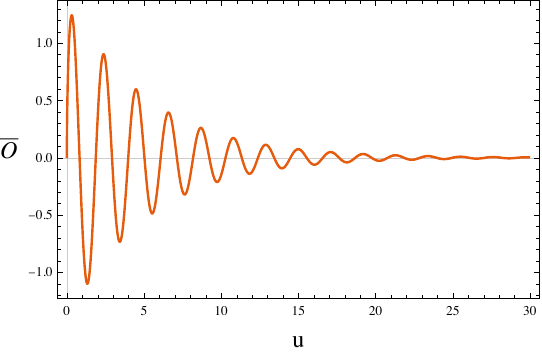}}
\subfigure[]{\includegraphics[scale=0.7]{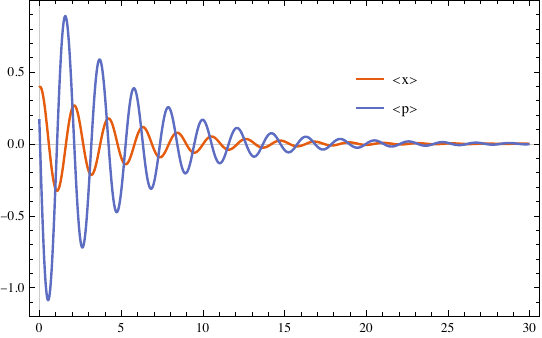}}
\caption{(a)$\bar{O}(u)= N^{-1} O(u)$ eventually decays to zero. (b) Evolution of $\braket{x}$ and $\braket{p}$ which also decay to zero. Both imply decoupling. We have chosen $\omega =3.2 N^2$, $\gamma = -0.3 N$ (with $N$ given by \eqref{Eq:N}) and $\Delta=\frac{5}{4}$.}
\label{Fig:osc_quant}
\end{figure}

The evolution of $\braket{x(u)}$, $\braket{p(u)}$ and $O(u)$ are shown in Fig.\ref{Fig:osc_quant}. We find that all of these eventually vanish confirming dynamical decoupling of the oscillator from the black hole. This can be compared with the dynamics in the classical limit of the oscillator described in Appendix \ref{Sec:AppendixA}. In the classical limit we also find the dynamical decoupling of the oscillator with the classical position $x$, momentum $p$ and $O$ asymptotically decaying to zero. However, unlike the quantum case, the classical oscillator looses all of its energy to the black hole and settles down to equilibrium. Note that all these results don't qualitatively depend on the sign of $\gamma$, the coupling constant for the coupling between the oscillator and the black hole.

Remarkably, the late time state of the quantum oscillator is not an energy eigenstate, but a quantum trajectory which can be determined analytically just from the decoupling when the initial state is in a qubit subspace as in \eqref{Eq:psi-in}. Since there is decoupling at late time, the oscillator should satisfy the free Schr\"{o}dinger equation implying that its intrinsic energy $E_{osc}$ as defined in \eqref{Eq:Eosc} should indeed go to a constant, and the time-dependent late-time state \eqref{Eq:psi-sol} should evolve to
 \begin{equation}
      \ket{\psi_L(u;x_L(u),p_L(u)} = \cos\theta e^{i\phi} \ket{\phi_0(u;x_L(u),p_L(u))}+ \sin\theta  \ket{\phi_1(u;x_L(u),p_L(u))}
        \label{Eq:fin_state_osc}
 \end{equation}
 given by functions $x_L(u)$ and $p_L(u)$ (instead of  $x_p(u)$ and $p_p(u)$) which obey the free classical equations 
\begin{equation}\label{Eq:xL-pL-eq}
    \ddot{x_L}(u) + \omega^2 x_L(u) = 0, \quad p_L (u) = \dot{x}_L(u).
\end{equation}
We recall that $\ket{\phi_0}$ and $\ket{\phi_1}$  are given by \eqref{Eq:phin-def} and \eqref{Eq:Fn-def} as functionals of $x_L(u)$ and $p_L(u)$. It follows that the final quantum trajectory of the oscillator is determined simply by the solution of \eqref{Eq:xL-pL-eq}, and therefore is labelled just by two parameters $A_0$ and $\xi_0$ (amplitude and phase) which parametrizes the general solution of \eqref{Eq:xL-pL-eq} via the form
\begin{equation}\label{Eq:A-chi-def}
   x_L(u) = A_0 \cos(\omega u + \xi_0).
\end{equation}
Asymptotically, we obtain
\begin{eqnarray}
    \braket{x(u)} &=& \braket{\psi_L (u)| \hat{x}| \psi_L(u)} \nonumber\\&=&x_L(u) +\sqrt{\frac{\hbar}{2 \omega}}\sin(2\theta)\, \cos(\omega u+\phi)
\end{eqnarray}
However, decoupling implies $\braket{x(u)} =0$, which further implies that the parameters $A_0$ and $\xi_0$ labelling $x_L(u)$ are
\begin{equation}\label{Eq:A-chi-det}
        A_0 = - \sqrt{\frac{\hbar}{2 \omega}}\sin(2\theta)\quad  \xi_0 = \phi,
\end{equation}
and thus the decoupling condition determines the final quantum trajectory completely. Furthermore, we note that \textit{the final quantum trajectory encodes the initial state \eqref{Eq:psi-in} completely in terms of the Bloch angles $\theta$ and $\phi$}. The encoding is however non-isometric. We also note that in the limit $\hbar\rightarrow 0$, we get $A_0\rightarrow 0$ implying that the classical limit is smooth (also note that $\vert A_0\sqrt{2\omega}\vert < \vert \hbar \vert$, so the quantum effects are small). Furthermore, since $\omega \sim \mathcal{O}(N^2)$ as discussed in the previous section, the amplitude of the residual oscillations is $\mathcal{O}(N^{-1})$ as evident from \eqref{Eq:A-chi-det}. The residual energy is of the order of $A_0^2\omega^2 \sim \hbar\omega \sim\mathcal{O}(\hbar N^2)$, and also goes to zero in the classical limit in which $\hbar \rightarrow 0$.

We can also compute the transfer of energy from the oscillator to the black hole analytically. From \eqref{Eq:fin_state_osc}, \eqref{Eq:A-chi-def} and \eqref{Eq:A-chi-det}, we can compute the final energy of the oscillator to be 
\begin{equation}
    \begin{aligned}
     E_{f,osc}&= \frac{1}{2m} \braket{p^2} + \frac{1}{2} m\omega^2 \braket{x^2} \\
     &= \frac{\hbar\omega}{2}\left(\cos^2 (\theta) +3 \sin^2(\theta) -\frac{1}{2}\sin^2 (2\theta)) \right)
    \end{aligned}
\end{equation}
The initial state \eqref{Eq:psi-in} gives the initial energy of the oscillator $E_{i,osc}$. Since the interaction energy goes to zero and the total energy is conserved (see the previous section), the energy transferred to the black hole is then
\begin{equation}\label{Eq:Entransfer}
    \Delta E = E_{i,osc}-E_{f,osc} = \frac{\hbar \omega}{4}\sin^2(2\theta),
\end{equation}
which is positive. Note that the maximum transfer of energy to the black hole is $1/2$ of the ground state energy of the oscillator, and this happens for $\theta = \pi/4$, when the initial energy of the oscillator is twice of that of the ground state. The Bloch angle $\theta$ is restricted between $0$ and $\pi$. The transfer of energy is zero when $\theta = 0, \pi/2, \pi$. In all these cases, the initial state \eqref{Eq:psi-in} is an energy eigenstate of the decoupled oscillator, i.e. it is either $\ket{0}$ or $\ket{1}$ up to a phase. In such cases, $\braket{x(u)} =0$ for all time, and so the oscillator never couples to the black hole if initially $O(u) =0$ as we have considered here.

Finally, we study how the initial (quantum) state of the oscillator is completely encoded into the black hole ringdown. As seen in the red curve in Fig. \ref{Fig:energy_quant_osc}, at late times the mass of the black hole is well described by an oscillating function with an exponentially decaying amplitude as
\begin{equation}
    M(u) = M_f - e^{-\tau u} \left(B \sin(\Omega u + \xi) + C \right),
\end{equation}
where $M_f$ is the final value of the mass and the amplitude of the ringdown is defined to be $A = B + C$. We also note that the phase $\xi$ is defined with respect to the time after which we start collecting the late time data. The decay rate of the ringdown is $\tau^{-1}$ and $\Omega$ is the frequency of the decaying oscillations. See Appendix \ref{Sec:AppendixC} for details on data fitting. Note that this ringdown is not the usual quasi-normal mode response of a black hole that is obtained via linearization, but it is a consequence of the full non-linear equations. By fitting the late time data we can numerically subtract out the exponential decay and do a discrete Fourier transform to extract the frequency $\Omega$, the amplitude $A$, the phase $\chi$ and the decay time scale $\tau$.
\begin{figure}
\centering     
\subfigure[]{\includegraphics[scale=0.7]{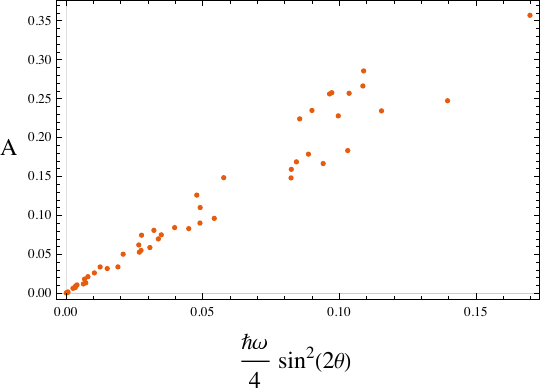}}
\subfigure[]{\includegraphics[scale=0.7]{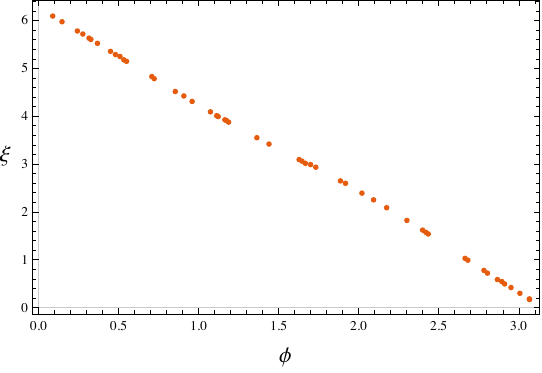}}
\caption{(a) The amplitude of the ringdown $A$ vs $\hbar \omega \rho_{01} \rho_{01}^*$ for different initial conditions. Note $\rho_{01} \rho_{01}^*= \frac{1}{4} \sin^2(2\theta)$ generally. (b) The phase of the ringdown $\xi$ vs  $\phi$ (note $\phi = \arctan\left(\frac{\rm Im \rho_{01}}{\rm Re \rho_{01}}\right)$ generally) for different initial conditions. We have chosen $\omega =3.2 N^2$, $\gamma = -0.3 N$ (with $N$ given by \eqref{Eq:N}) and $\Delta=\frac{5}{4}$. The outliers are essentially due to the errors involved in computing numerical Fourier transforms of the data.  }
\label{Fig:encode_quant_osc}
\end{figure}

The initial state \eqref{Eq:psi-in} is completely encoded into the late time ringdown parameters, $A$, $\chi$ and $\tau$. The amplitude $A$ and phase $\xi$ encode the Bloch sphere angles $\theta$ and $\phi$ respectively, as shown in Fig.\ref{Fig:encode_quant_osc}, and  therefore the initial state \eqref{Eq:psi-in} is completely and non-isometrically encoded into the black hole ringdown. It is interesting to compare \eqref{Eq:A-chi-det} with the plots shown in Fig.\ref{Fig:encode_quant_osc}. We notice that the amplitude of the black hole ringdown scales quadratically with $\sin(2\theta)$ instead of linearly like $A_0$, while the phase $\xi$ depends linearly on $\phi$ like $\chi$, but with a negative slope. Also the amplitude of the ringdown is proportional to the transfer of energy of the oscillator to the black hole mass \eqref{Eq:Entransfer}.

\begin{figure}
\centering     
\includegraphics[scale=0.7]{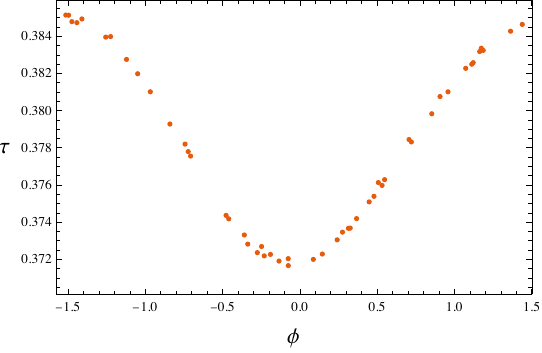}
\caption{The decay exponent also encodes the angle $\phi$ on the Bloch sphere.}
\label{Fig:mixed_encode_osc}
\end{figure}

As shown in Fig. \ref{Fig:mixed_encode_osc}, the decay time scale $\tau$, which has a narrow distribution, also encodes the Bloch angle $\phi$ (which lies between $-\pi$ and $\pi$) up to an overall sign. We note From Fig. \ref{Fig:mixed_encode_osc} that $\Delta \tau/\tau_{av} < 10^{-1}$, where $\Delta \tau$ is the range of variation of $\tau$ and $\tau_{av}$ is the average value of $\tau$ over the range of variation. If we specify $\omega/N^2$ and $\gamma/N$ (which are $\mathcal{O}(N^0)$) with $N$ defined via \eqref{Eq:N}, then the equation of motion for the black hole mass \eqref{Eq:WI2} has no explicit $N$ dependence (see the discussion in the previous section). Therefore, $\tau$ of the black hole ringdown does not depend on $N$ explicitly, but depends on the parameters $\omega/N^2$ and $\gamma/N$ along with the Bloch angle $\phi$ of the input state. 

It is also natural to speculate that the frequency of the ringdown $\Omega$ encodes some of the information of the initial state. However, it is hard to determine $\Omega$ (which has a very narrow range of variation like $\tau$) to the desired accuracy from the Fourier transform needed to correlate with the data of the initial state. Like $\tau$, $\Omega$ also does not depend on $N$ explicitly but on the parameters $\omega/N^2$ and $\gamma/N$ along with certain parameters of the input state.

We can repeat the above analysis for a mixed initial state. Here again we assume that the initial state of the oscillator is restricted to the subspace which is the span of $\ket{0}, \ket{1}$. We parametrize the initial density matrix as
\begin{align}
    \rho(u=0) = \rho_{00}\ket{0}\bra{0} + \rho_{01} \ket{0}\bra{1} + \rho_{01}^*\ket{1}\bra{0} +(1-\rho_{00})\ket{1}\bra{1},
\end{align}
where $\rho_{00}$ is real and $\rho_{01}^*$ is the complex conjugate of $\rho_{01}$. Using the same arguments as for the pure state we find that the following state is a solution to the Liouville equation for the time evolution of the density matrix
\begin{equation}
    \rho(u) = \rho_{00}\ket{\phi_0(u)}\bra{\phi_0(u)} + \rho_{01} \ket{\phi_0(u)}\bra{\phi_1(u)} + \rho_{01}^*\ket{\phi_1(u)}\bra{\phi_0(u)} +(1-\rho_{00})\ket{\phi_1(u)}\bra{\phi_1(u)},
\end{equation}
where $\ket{\phi_{0,1}}$ depend on $x_p,p_p$ as in \eqref{Eq:phin-def} and \eqref{Eq:Fn-def}. The qualitative behaviour of the energies and other observables is found to be the same as shown in Fig.\ref{Fig:energy_quant_osc} and Fig.\ref{Fig:osc_quant} such that the oscillator and the black hole eventually decouple with $\braket{x(u)} \to 0$ and $O(u)\to 0$. We also find that the transfer of energy from the oscillator to the black hole is
\begin{equation}
    \Delta E =  \hbar \omega \rho_{01} \rho_{01}^*.
\end{equation}
which agrees with \eqref{Eq:Entransfer} (generally $\rho_{01} \rho_{01}^*= (1/4) \sin^2(2\theta)$).

The initial state can be decoded from the black hole ringdown as in the case of the pure state. The decay rate $\tau$ and phase of the ringdown $\xi$ encodes the Bloch sphere angle $\phi$ (note $\phi = \arctan\left(\frac{\rm Im \rho_{01}}{\rm Re \rho_{01}}\right)$ generally) as shown in Fig.\ref{Fig:encode_quant_osc} and \ref{Fig:mixed_encode_osc}. The amplitude encodes the Bloch angle $\theta$  as shown in Fig.\ref{Fig:encode_quant_osc}.  Thus for any arbitrary mixed initial state of the oscillator within a qubit subspace, we can decode both ${\rm Re}(\rho_{01})$ and ${\rm Im}(\rho_{01})$ from the late time ringdown of the black hole mass. The only missing piece of information is the purity of the state, which is not well correlated with the amplitude, phase or decay exponent of the ringdown. We suspect that the purity of the state is correlated with the frequency $\Omega$ of the ringdown, but we need to improve our numerical accuracy for a confirmation. Moreover, as seen from the final state of the oscillator in \eqref{Eq:fin_state_osc}, the oscillator retains complete information about its initial state (the angles $\theta$ and $\phi$). 

For a general initial state which is a superposition of more than two Fock states, the evolution can be solved readily by the method of invariants discussed here, and we find that the decoupling of the oscillator from the black hole is a general phenomenon. Since the evolution of the oscillator is unitary (more precisely it is a non-linear unitary mean field channel) in the semi-classical large $N$ approximation, the final state retains complete information of the initial state non-isometrically, although it cannot be obtained simply from the decoupling condition $\braket{x(u)}\to 0$ as we have done for the initial state \eqref{Eq:psi-in}.  However, the dominant decay mode of the black hole ringdown with its decay rate, amplitude and phase is of course insufficient to encode the initial state. We need to extract the sub-asymptotics of the black hole ringdown for which more analytic understanding of the system and also higher numerical accuracy will be needed. We plan to investigate this in the future.


\subsection{Entropy for the single throat model}
\begin{figure}
\centering     
\includegraphics[scale=0.8]{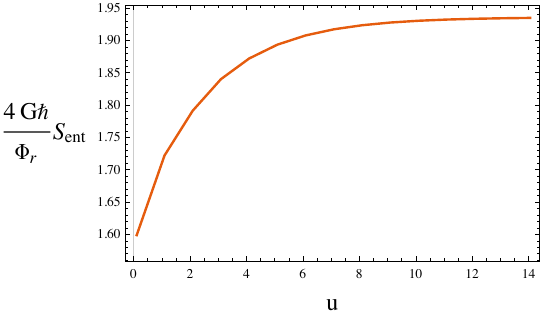}
\caption{The entropy of the quantum dot state dual to the $AdS_2$ throat for the single throat model is monotonically increasing.}
\label{Fig:entropy_qubit}
\end{figure}

\begin{figure}
\centering     
\includegraphics[scale=0.7]{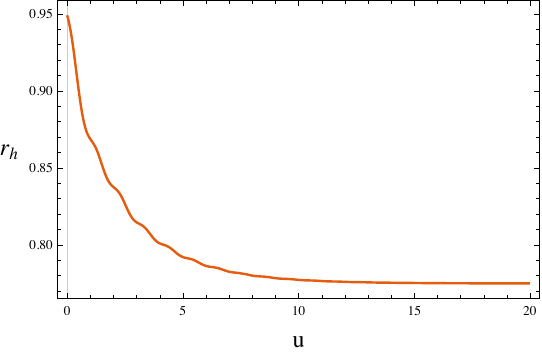}
\caption{The position of the dynamical event horizon $r_h(u)$ with $l=1$. The horizon smoothly interpolates between the initial and final thermal horizons for a particular initial condition where the final mass is 1.66}
\label{Fig:rhplot}
\end{figure}

In JT gravity, the entanglement entropy of the state of the dual theory can be holographically computed using the Ryu-Takayanagi formula \cite{Ryu:2006bv,Hubeny:2007xt} as 
\begin{equation}
    S_{\rm ent} = \frac{\Phi(r_h)}{4 G \hbar},
\end{equation}
where $r_h$ is the location of the event horizon, which smoothly interpolates between the initial and final thermal horizons. This can be found as follows. We map the metric \eqref{Eq:metricEF} to  the metric for a black hole with fixed mass $M=\frac{1}{l^2}$
\begin{equation}
    ds^2 = -2 \frac{l^2}{R^2} dT dR - \left(\frac{l^2}{R^2} -  1 \right) dT^2,
\end{equation}
via
\begin{align}
    &T = T(u),\quad R =\frac{T'(u) r}{1-\frac{T''(u)}{T'(u)}r},\\
    &-2 Sch(T(u),u) +T'(u)^2= M(u). 
\end{align}
The horizon for the $M=\frac{1}{l^2}$ black hole is simply $R_h = \frac{1}{\sqrt{M}} = l$. We can invert the diffeomorphism above to find a dynamical horizon for the mass $M(u)$ black hole, which is \cite{Joshi:2019wgi}
\begin{equation}
    r_h(u) = \frac{l}{T'(u) \left(1+\frac{T''(u)}{T'(u)^2}\right)}.
\end{equation}
This horizon smoothly interpolates between the initial and final thermal horizons as shown in Fig.\ref{Fig:rhplot}. We then use the solution for the dilaton \eqref{Eq:dilaton_soln} to find the entropy. As shown in Fig. \ref{Fig:entropy_qubit}, the entropy increases monotonically. The evolution of the oscillator is unitary due to the semi-classical mean field approximation and the bulk scalar is classical, therefore there are no other time dependent entropies that we need to take into account. The monotonicity of the entropy of the quantum dot thus demonstrates the consistency of this model with the second law of thermodynamics. 

The monotonic entropy growth reflects the irreversible loss of information in the semi-classical geometry of the throat as the black hole ringdown encodes only a transitory copy of the information of the initial state. This information is however encoded into the Hawking radiation. As described in the next section, the loss of energy via Hawking radiation is proportional to $M(u)$ itself, as for instance when the Hawking quanta is made of conformal fields in the black hole background. Therefore, a continuous fine grained measurement of the rate of flow of energy via the Hawking quanta (if sufficiently weak) can decode the initial state of the qubit. 

\section{Hawking radiation, grand decoupling and the Hayden-Preskill protocol}
\label{Sec:Hawkingrad}

\subsection{The evaporating microstates and the complementarity principle}
The study of the single throat model in the semi-classical approximation reported above reveals that decoupling of the remnant infalling matter in the horizon hair from the interior of the black hole is the key to the information replication after the irreversible transfer of energy to the black hole interior. During the decoupling, the time-dependent state of the remnant quantum matter encodes non-isometrically the initial state of the matter while the ringdown of the throat encodes another non-isometric transitory copy of the infalling matter which is transferred to Hawking radiation. Therefore, it is necessary to investigate whether the decoupling generalizes to the full microstate model during Hawking evaporation. 

Since the decoupling can play a role for an elegant encoding of the initial conditions of the black hole formation itself into Hawking radiation, we study the evaporating microstate first without infalling matter before studying information mirroring of infalling matter during the evaporation process in the next subsection.


Hawking evaporation can be readily incorporated into our microstate model defined in Sec. \ref{Sec:microstate} simply by (i) introducing conformal matter in each throat with same central charge $c$ and (ii) coupling each throat to a non-gravitating bath which collects the Hawking quanta following \cite{AEMM} (see Appendix \ref{Sec:AppendixD} for a brief review).  This leads to a modification of the equations for the microstate model given by \eqref{Eq:LatticeEoms1} and \eqref{Eq:LatticeEoms2} as follows:
\begin{align}
\dot{M}_i &= - \la(\bQ_{i-1}+ \bQ_{i+1}-2 \bQ_i)\dt \dot{\bq}_i - \frac{c}{24 \pi} M_{i},\label{Eq:LatticeEoms4}\\
\ddot{\bq}_i &=\frac{1}{\si^2}( \bq_{i-1} + \bq_{i+1} - 2 \bq_i) +  \frac{1}{ \la^2}(\bQ_{i-1}+ \bQ_{i+1}-2 \bQ_i)\label{Eq:LatticeEoms5}.
\end{align}

\begin{figure}
\centering
\includegraphics[scale=0.7]{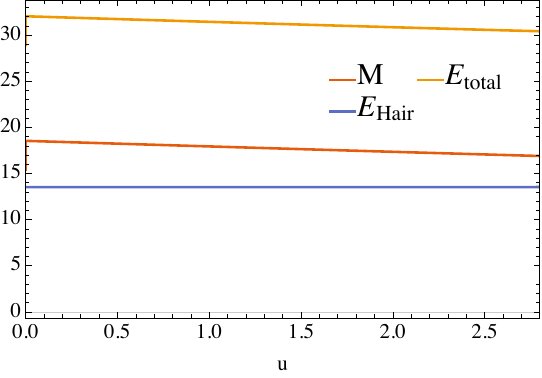}\caption{Plot of the different energies for our model coupled to a CFT bath. Where $E_{\rm Hair} = E_{\bq}$, $M = \sum_i M_i$. A decay of the total mass and total energy due to Hawking radiation can be seen.}\label{Fig:PageEnergy}
\end{figure}

\begin{figure}
\centering
\includegraphics[scale=0.7]{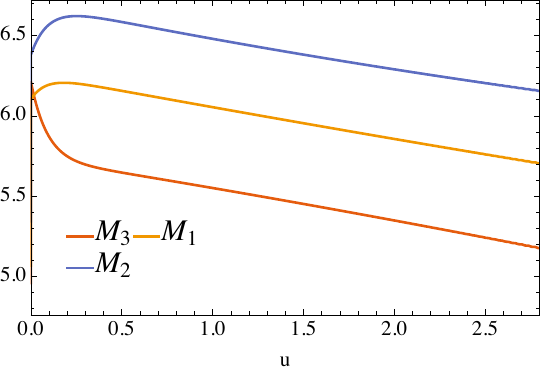}\caption{Plot of the mass of each site in the 3 site model.}\label{Fig:PageMass}
\end{figure}

The masses of the individual throats decay as shown in Fig.\ref{Fig:PageMass}.
The typical evolution of energies, namely the total ADM mass $M = \sum_i M_i$, the total energy of the hair $E_{\bq}$ and their sum $E$ (see \eqref{Eq:Enmicro1} - \eqref{Eq:Enmicro3} for definitions), for a lattice with $3$ sites arranged on a circle, after open boundary conditions are imposed at $u=0$ at the interface of the bath and the throat to allow Hawking radiation to escape to the bath, is shown in Fig.\ref{Fig:PageEnergy}. Recall that $E$ is conserved in absence of Hawking radiation. Remarkably, $E_{\bq}$ changes extremely slowly even compared to the decay of $M$ during Hawking evaporation. The evolution of the individual masses of the three throats is shown in Fig. \ref{Fig:PageMass}. These are non-monotonic functions of time, but they decay at late time.


\begin{figure}
\centering
\includegraphics[scale=0.7]{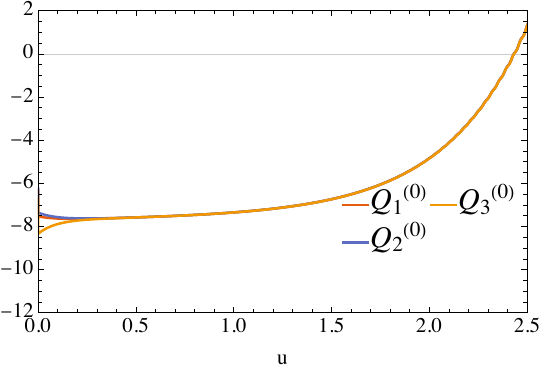}\caption{Evolution of $\Q_i^0$. Note they converge asymptotically.}\label{Fig:PageQ0}
\end{figure}

The change in boundary condition at $u=0$ introduces a null shock at each throat of any pre-existing microstate characterized in Sec. \ref{Sec:microstate}. (This shock at $u=0$ has been excised from Figs.\ref{Fig:PageEnergy} and \ref{Fig:PageMass}.) We recall that one of the components of $\bQ_i$ (the $SL(2,R)$ charges of the throats) should be homogeneous in any microstate, and this component has been chosen to be the zeroth component by exploiting the global $SL(2,R)$ symmetry. As in the case of null shocks studied in Section \ref{Sec:microstate}, the change in boundary condition leads to a jump in all components of $\bQ_i$ in all the throats. As shown in Fig.\ref{Fig:PageQ0}, the $\bQ_i^{0}$ components homogenize exactly as discussed in the case of non-evaporating microstates subjected to null shocks. However, $\bQ^{0}_i$ doesn't settle down to a constant after homogenization. Furthermore, the time scale of this homogenization is very different from the relaxation time scale discussed earlier and plotted in Fig. \ref{Fig:relaxtime}. Remarkably, the other components of the $SL(2,R)$ charges of the throats, i.e. $\bQ_i^\pm$, also homogenize over the same time-scale. The upshot is that the $SL(2,R)$ charges of all the throats asymptote to a common null vector as the masses of all the throats decay. 

The consequence of homogenization of all components of $\bQ_i$ is that all components of the hair charges, namely $\bq_i$, decouple from the throats and satisfy the free discretized Klein-Gordon equations as evident from \eqref{Eq:LatticeEoms5}. Furthermore, homogenization  implies that at late time all the masses of the throats should decay independently at the same rate given by the central charge $c$ as evident from \eqref{Eq:LatticeEoms4}. Therefore, the homogenization time-scale is a \textit{grand} decoupling time-scale at which all throats decouple from each other, and all components of the hair charges also decouple from all the throats.

\begin{figure}
\centering
\includegraphics[scale=0.7]{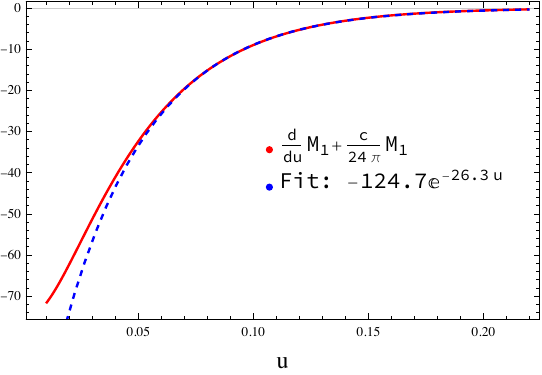}\caption{Evolution of $\dot{M}_i+\frac{c}{24 \pi} M_{i}$ for $i=1$ in a 5 site model. The dashed line is a fit of the data to an exponentially decaying function. The fit has an adjusted R squared value of 0.99997 for data in the range $u \in [0.07,0.2]$.}\label{Fig:decouplefit}
\end{figure}

To extract the decoupling time scale, we consider $$D_i \equiv \dot{M}_i+\frac{c}{24 \pi} M_{i} = - \la(\bQ_{i-1}+ \bQ_{i+1}-2 \bQ_i)\dt \dot{\bq}_i$$where the last equality is a consequence of the equation of motion \eqref{Eq:LatticeEoms4}. As shown in Fig.\ref{Fig:decouplefit}, the modulus of $D_i$ for each throat decays to zero due to the homogenization of the $SL(2,R)$ charges. We fit this to an exponentially decaying function $$D_i \sim d_i \exp(-u/t_{di})$$and extract the timescale $t_{di}$ which does depend on the particular throat. We can define the grand decoupling time $t_d$ for the particular initial microstate as the average of $t_{di}$ over the $n$-throats, i.e. $t_d = (1/n)\sum_{i} t_{di}$. 

\begin{figure}
\centering
\includegraphics[scale=0.7]{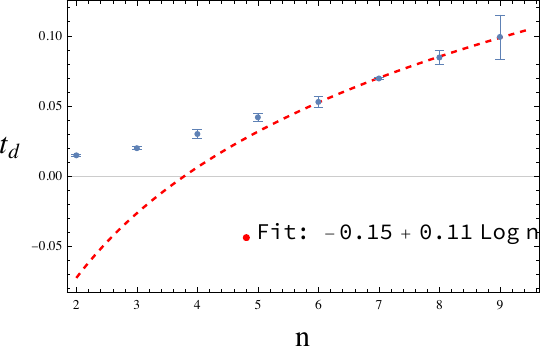}\caption{Variation of the decoupling time with the number of sites $n$. For large values of $n$ the data is well described by a logarithm (red dashed line). The solid dots indicate the mean of the decoupling time over various initial conditions and the error bars indicate the standard deviation. The fit has an adjusted R squared value of 0.999909 for data in the range $n\in [6,9]$.}\label{Fig:granddecouple}
\end{figure}

A plot of $t_d$ as a function of $n$, the number of throats, with the average mass per throat, i.e. $M/n$ held fixed is shown in Fig. \ref{Fig:granddecouple}. Recall that the average mass per throat is a proxy for the temperature. Unlike the relaxation time $t_r$ studied in Section \ref{Sec:microstate} (and plotted in \ref{Fig:relaxtime}), the decoupling time does not saturate to a constant as we increase $n$ and is therefore not proportional to the inverse temperature. In fact, as shown in Fig. \ref{Fig:granddecouple}, the decoupling time averaged over several initial microstates, grows as $\log n$ at larger values of $n$ when the average mass per site is fixed. 

We note that the entropy ($e^{S}$ giving the number of microstates) scales linearly with $n$ for large $n$ and fixed (average) mass. Since we need to specify a few independent variables at each lattice site subject to a few global constraints including the mass, the number of microstates scales exponentially with $n$ and therefore the entropy should scale linearly with the number of sites when the latter is large. Therefore $t_d$ scales as $\log S$ when the average mass per site is held fixed. It then seems that $t_d$ could be identified with the scrambling time scale which also scales as $\log S$. Nevertheless, we need to investigate $t_d$ for a larger number of lattice sites and our current numerical methods limits us from studying evolution of the radiating microstates for $n> 9$ over time scales commensurate with the decoupling time scale (see \ref{Sec:AppendixB} for a discussion on the numerical challenges). In the following section, we will argue why $t_d$ can be indeed identified with the scrambling time scale.

At first glance the decoupling would seem paradoxical as the interior homogenizes and all information is lost therein although the decoupled hair still retains information. However, as we have seen earlier, the details of the ringdown of each throat at late time after decoupling, as for instance the distribution of $t_{di}$, can capture the initial information and elegantly transfer the same to the Hawking radiation. As mentioned earlier, numerically it is harder to extract more sub-dominant decay scales and frequencies reliably, so we need to increase the numerical accuracy to a large extent before we can study this better.


\subsection{Information mirroring, Hayden-Prekill protocol and the complementarity principle}

The basic argument for quantum information mirroring of an old black hole is essentially from quantum statistics \cite{Hayden_2007}. Consider a black hole microstate $B$ and its emitted Hawking quanta $E$. A quantum diary $D$ entangled with a reference system $S$ (used just as a tool for the argument) interacts with the microstate $B$ so that it transits to another $B'$ emitting Hawking quanta $R$ via a \textit{random} unitary operator $U$. If $R$ has a small number of bits compared to $B$ (and therefore also $E$ after Page time), then we can show that $S$ and $B'$ have no mutual entanglement. Therefore, the information of the diary $D$ is in $R\cup E$. The information of any infalling bit should therefore come out of the black hole after the scrambling time over which the Hawking quanta escapes the black hole, i.e. the time-scale of the unitary gate $U$. A microstate model of a quantum black hole should not only reproduce the Page curve \cite{Page_1993,Page_2013} but also quantum information mirroring.

For quantum information mirroring to be consistent with black hole complementarity principle, the decoding of the diary $D$ from $R\cup E$ should be possible without knowledge of the microstate $B$ prior to the infall and also the microstate $B'$ that results from the scrambling process \cite{Kibe:2021gtw}. The interior of the black hole is also encoded into the Hawking quanta, but it has been argued that it should take much longer time (that scales with the exponent of the black hole entropy at the time of emission) to decode this (by first distilling the part of $E$ which purifies $R$) \cite{Harlow:2013tf,Aaronson:2016vto,Kim:2020cds}, as otherwise we will not be able to operationally avoid the AMPS paradox that arises from assuming that all postulates of the black hole complementarity holds for semiclassical observables . Furthermore, it should be also possible to decode the mirrored information without requiring any detailed knowledge of the pre-emitted radiation $E$ which is far away from the newly emitted quanta $R$ if an effective description is valid and the horizon is smooth as assumed by the black hole complementarity principle \cite{Kibe:2021gtw}.  

Quantum information mirroring has been studied previously in our microstate model without incorporating Hawking radiation \cite{PhysRevD.102.086008}. Even in the absence of Hawking radiation, the argument for quantum information mirroring extends to the hair degrees of freedom which decouple from the interior of the microstate, and therefore it is a very strong consistency test of the model. To test information mirroring, classical information was encoded into a sequence of null shocks injected into the throats, so that the information is in a time-ordered list with each entry denoting the location of the throat which is shocked. As described in Sec. \ref{Sec:microstate}, the microstate quickly transits to another after the shocks (the dynamics has characteristics of pseudorandomness) while the $\bq_i^0$ component of the hair decouple from the interior and satisfy a discretized Klein-Gordon equation. We recall that we can always set the decoupled component to be the zeroth one by using the global $SL(2,R)$ symmetry, but this particular component also agrees with the pre-exisiting zero mode of the hair radiation (the monopole term discussed in Sec. \ref{Sec:microstate}). The other components of the hair, namely $\bq_i^\pm$ get locked to the final configuration of the $SL(2,R)$ charges of the throats. 

\begin{figure}
\centering
\subfigure[Two non-neighboring sites are shocked. The dark blue site is shocked first and the light blue one later. The site shocked first has the highest phase difference between the two positive normal modes and the site shocked later has the smallest phase difference.]{\includegraphics[width = 2in]{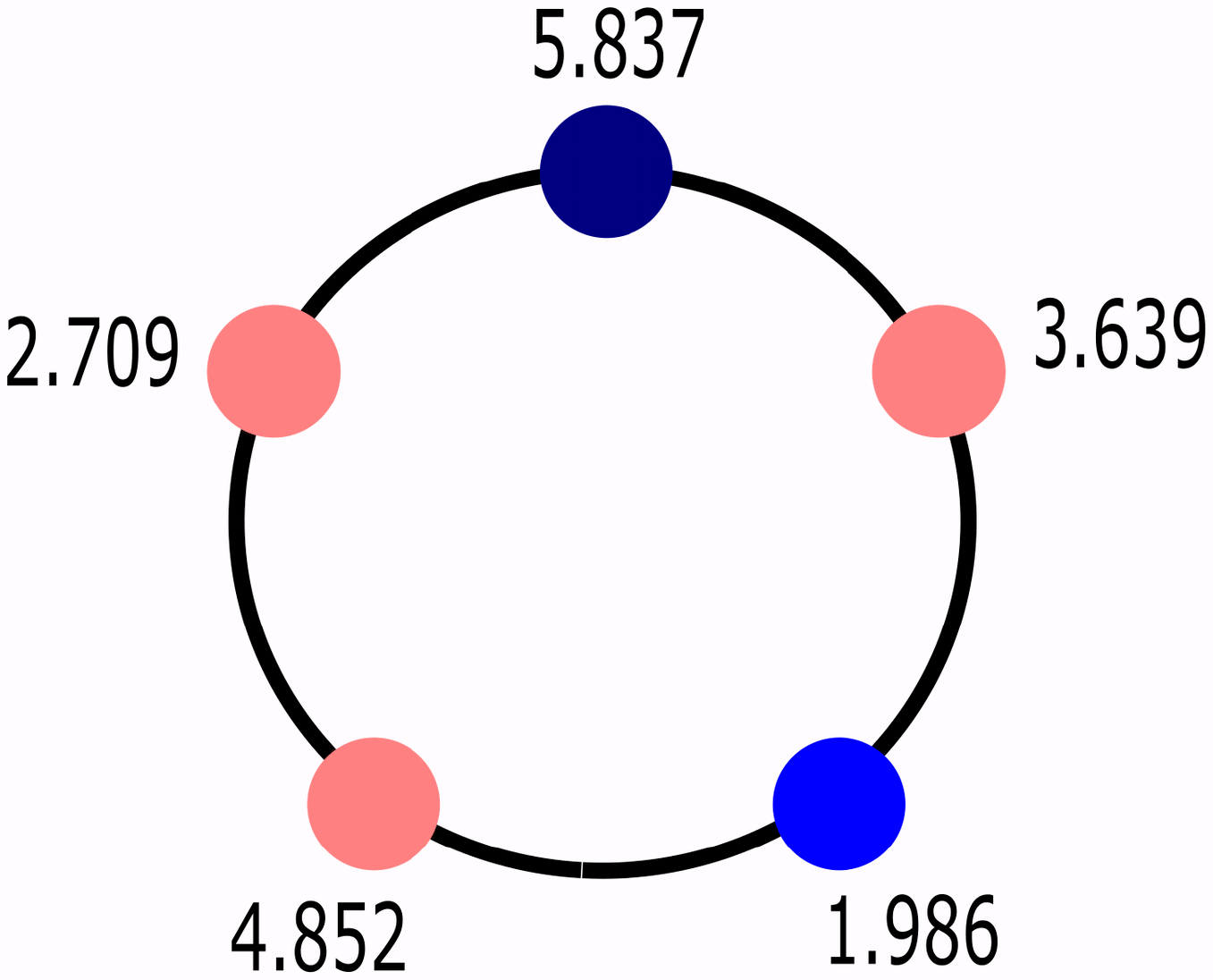}}\\
\subfigure[Two neighboring sites are shocked. The dark blue site is shocked first and the light blue one later. The site shocked first has the highest phase difference between the two normal modes and the site shocked later has the smallest phase difference.]{\includegraphics[width = 2in]{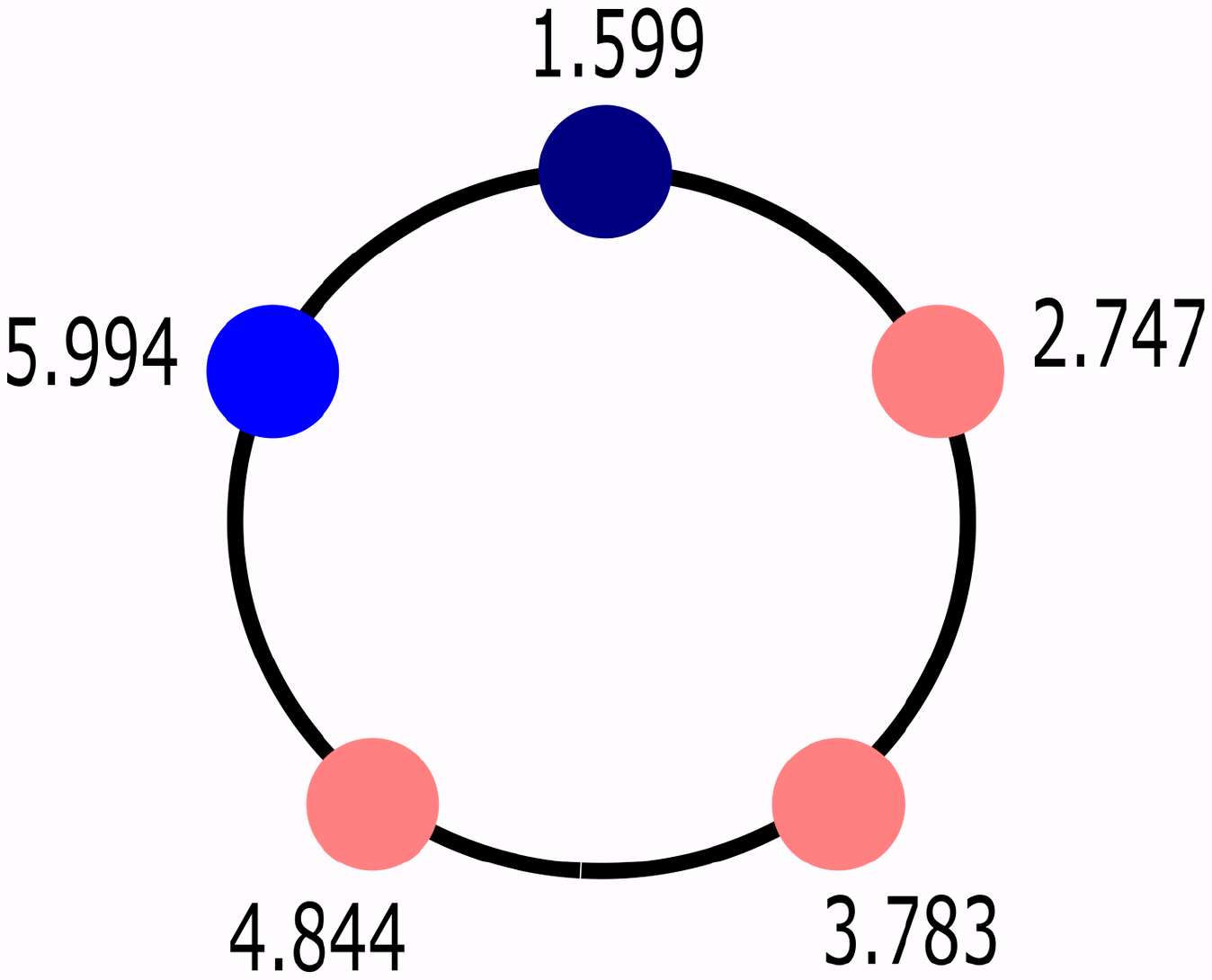}}\\
\subfigure[A single site is shocked. The shocked site has the highest phase difference between the positive normal modes. ]{\includegraphics[width = 2in]{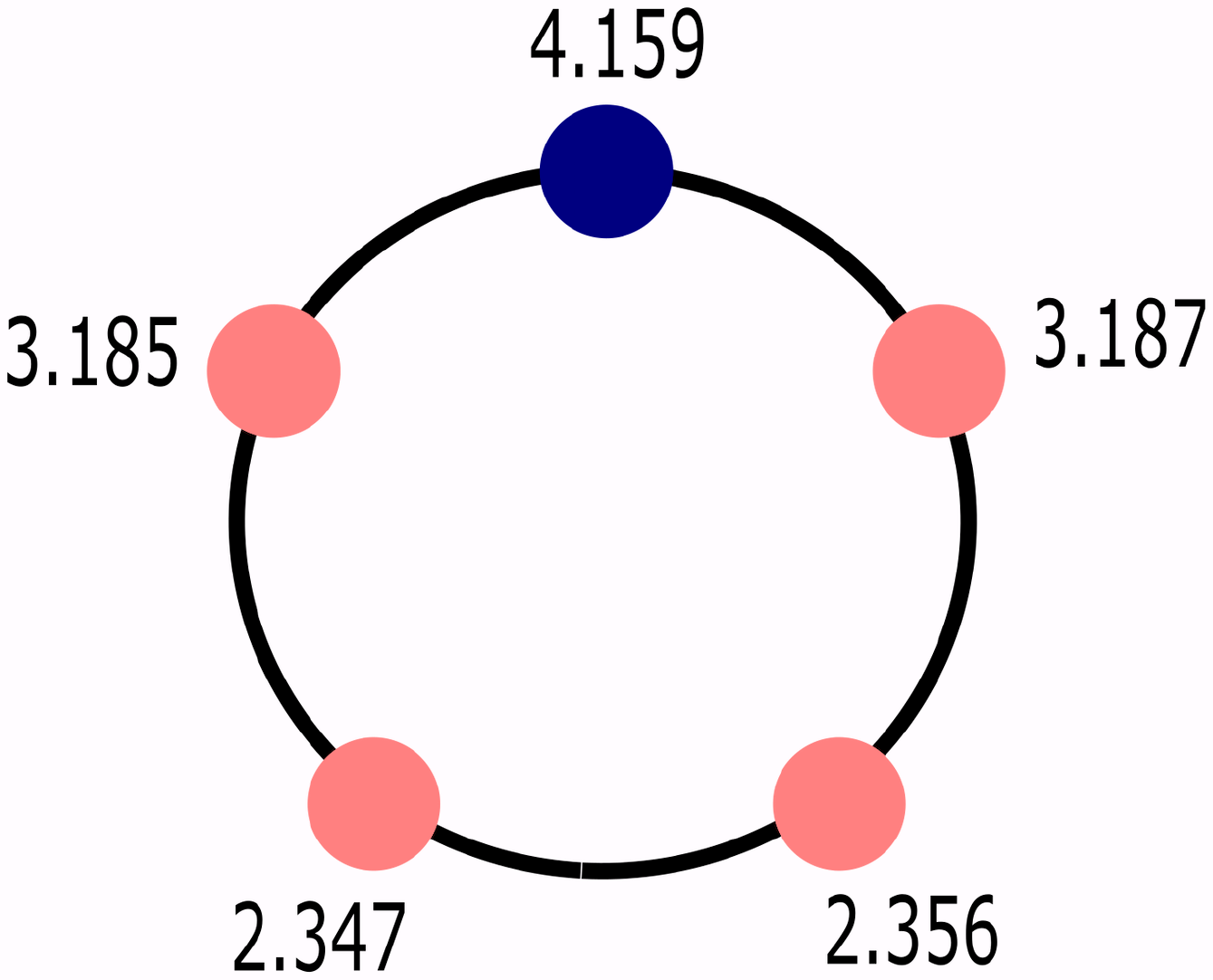}}
\caption{Hayden-Preskill decoding protocol for a 5 site chain model. This figure is from \cite{Kibe:2021gtw}. The numbers on each circle (representing a site) refer to phase difference of the two positive frequency decoupled normal mode oscillations.} \label{Fig:Mirroring}
\end{figure}

It was found that the time-ordered sequence of the location of the shocks could be decoded simply from the phase differences of the decoupled normal mode oscillations of $\bq_i^0$, and thus in consistency with the complementarity principle, requires no knowledge of the initial and final configuration of charges of throats describing the interior \cite{PhysRevD.102.086008}. Also, the only necessary knowledge of pre-existing radiation is essentially which component satisfies the free discretized Klein Gordon equation (the $\bQ_i^\pm$ components are inhomogeneous and time-independent after the decoupling, so do not satisfy the free equations). The decoding protocol is explicitly shown in Fig. \ref{Fig:Mirroring}. We readily note that for a single shock, the phase differences of the free decoupled normal mode oscillations are arranged symmetrically with the shocked throat having the highest phase difference. This is remarkable since both the initial and final microstate configurations are inhomogenoeous and thus asymmetric. The decoding of two shocks (where phase differences show similar symmetries) is described in Fig. \ref{Fig:Mirroring}.

We have presently investigated information mirroring for the evaporating microstate. Since in this simplified model we do not have the asymptotic region, and therefore no natural way to separate the early and late radiation, we can use the decoupled hair as proxy for the new radiation. As described in the previous subsection, \textit{all} components of the hair eventually decouple from the evaporating microstate, and thus satisfy the free discretized Klein-Gordon equation. The same decoding protocol Fig. \ref{Fig:Mirroring} described which applied to the $\bq_i^0$ components now applies to \textit{all} components of the hair radiation. 

For the decoding protocol, the only input needed from the bath collecting the Hawking radiation is to denote the time when decoupling has actually happened. We note the decoupling happens when all components of the charges of the throats, namely $\bQ_i$ homogenize, so that the rates of Hawking radiation (i.e. the loss of black hole mass) from each throat also homogenize. As evident from \eqref{Eq:LatticeEoms4}, when $(\bQ_{i-1}+ \bQ_{i+1}-2 \bQ_i)=0$ after homogenization, $\dot{M}_i/M_i = -c/(24\pi)$. Therefore, a continuous weak measurement of rate of energy transferred of Hawking quanta is necessary to determine the onset of mirroring of information. 

The mirroring of infalling information occurs only after the full decoupling. Therefore, the scrambling time should be identified with the decoupling time in our model. The $\log S$ growth of the decoupling time scale should then follow from that of the scrambling time at a fixed temperature (for which the average mass is a proxy) with $S$ being the entropy at the time of infall (since this couples back all degrees of freedom). Therefore, the $\log S$ growth of the decoupling time, which has been reported in the previous subsection for largish $n$ and fixed average mass, is a vindication of the model.

It should be possible to study the mirroring of quantum information in our model with more numerical accuracy and analytic understanding of the model. We discuss this in the following concluding section.

\subsection{An outlook: Replication, Mirroring and Entanglement}\label{Sec:outlook}

We have succeeded in establishing that decoupling of all the degrees of freedom happens generally in the microstate model, and furthermore this leads to quantum information mirroring. The throats decouple from each other and the hair decouple from the throats during the process of Hawking evaporation, and the mirroring of the infalling information occurs in the hair after the decoupling. We have observed that the decoupling time scales as the logarithm of the entropy like the scrambling time. Furthermore, the decoding protocol for the the mirrored information in hair does not require any knowledge of the interior state before the infall and the decoupling, and only very limited knowledge of the Hawking radiation. Here we have studied only mirroring of classical information encoded into the the time sequence of the locations of null shocks.

The two questions which naturally arise are as follows.
\begin{itemize}
    \item Can we see the replication of quantum information in the ringdown of the black hole and its subsequent transfer to the Hawking radiation, together with the mirroring of the information in the hair? 
    \item Can we also see how the black hole encodes its own initial state during the decoupling into the Hawking radiation and the hair?
\end{itemize}

The most fascinating aspect would be the replication and mirroring of information of two entangled qubits infalling into two different throats. It is natural to expect that the ringdown of the throats would be correlated after the decoupling, and unlike the case of the single throat model described in Sec. \ref{Sec:Replication}, we would need to study the correlated ringdown of all the throats (as for instance by seeing the decay of $\bQ_i\cdot \bQ_j$) to decode the information of the entangled infalling bits. This information would be transferred to the Hawking radiation. However, since the initial conditions would play a role in the encoding, we expect that the encoding into the Hawking radiation would be complex. To understand the mirroring of infalling information, we would need to quantize the $SL(2,R)$ hair also, which is feasible. 

\begin{figure}
\centering
\includegraphics[scale=0.35]{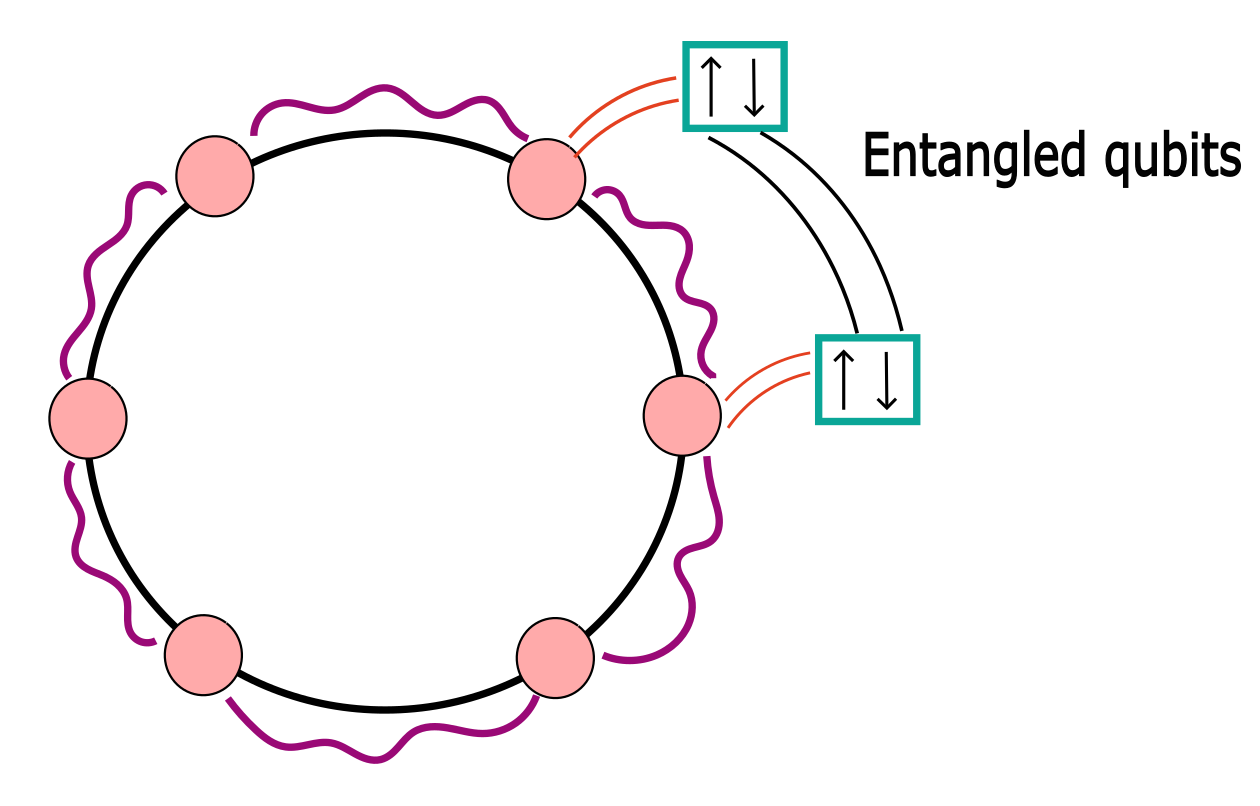}\caption{Entangled qubits can be coupled to two throats of the microstate model.}\label{Fig:entangledqubits}
\end{figure}

The general setup for this study is then as shown in Fig.\ref{Fig:entangledqubits}, where we need include classical bulk fields in the throat and couple them to quantum fields living on the discretized lattice as in Sec. \ref{Sec:Replication}. Additionally, we need to quantize the hair. This general setup can be solved numerically (rather semi-analytically since we know the quantum state up to some unknown functions as discussed in Sec. \ref{Sec:Replication}) using the methods described here. The challenge is to attain better numerical accuracy and analytic understanding. In Appendix \ref{Sec:AppendixB}, we have discussed the numerical challenges for simulating the full microstate model.

Another related question is to understand the entanglement wedges (islands) \cite{,AlmheiriQES,Almheiri2020Islands}. If we  understand which specific information of subspaces of the residual infalling quantum matter (which retains information of its own initial state in the unitary mean field approximation) is encoded into specific portions of the interior undergoing ringdown, then we can study the detailed encoding into Hawking radiation by finding out which subregions of the bath has entanglement wedges (islands) overlapping with the same portions in the interior which is encoding the infalling information. In this regard, the recent works \cite{Gyongyosi:2022vaf,Gyongyosi:2023sue} are relevant.

\section{Conclusions}
\label{Sec:conclusion}

In this paper, we have analyzed the previously proposed microstate model, and have further established that it behaves like a semi-classical black hole in terms of its energy absorption and relaxation properties. In particular, we have shown that the relaxation time is determined by the average mass per fragmented throat, which is a proxy for the temperature, in the thermodynamic limit. We have further shown that information replication of infalling quantum matter occurs in the simplified single throat model with one copy of the information retained in the residual time-dependent quantum state, and another copy encoded into the ringdown of the interior which transits from one microstate to another. The latter transitory copy is transferred to the Hawking radiation. The key to information replication is the decoupling of the quantum trajectories of the residual state in the hair and the black hole interior.

We have also studied the Hawking evaporation of the full microstate model and found that the fragmented throats decouple from each other and the hair decouples from the throats. This \textit{grand decoupling} is exactly what is needed to produce scrambled (and possibly multiple) copies of the information of the initial state of the black hole itself into the correlated ringdowns of the evaporating fragmented throats, and this decoupling also leads us to then explicitly understand the transfer of the information to the independent Hawking radiation from the throats. 

We have also shown that the decoupled hair mirrors infalling information such that the decoding can be done without the knowledge of the interior and only limited measurement of the Hawking radiation as can be argued to be necessitated by the complementarity principle. The decoupling time also scales with the logarithm of the entropy at fixed temperature, like the scrambling time, after which the mirroring should occur. This gives further evidence that decoupling is related to scrambling.

In the future, we hope to develop the numerical accuracy of our simulations and also further analytic understanding so that we can understand better how the correlated ringdowns of the decoupled throats encode (i) the initial conditions for the microstate, and (ii) replicated information of entangled qubits infalling into different throats. We further want to understand the detailed quantum information in the hair and see if all these generally point out to the emergence of the complementarity principle without paradoxes. The detailed general setup is discussed in Sec. \ref{Sec:outlook} which is numerically solvable using the methods discussed here.

Finally, we note that multiple replication of information can lead to new types of conceptual questions, which are as follows
\begin{itemize}
    \item What are limits on parallel processing of replicated copies of the information? This requires measurements done on each copy, but these can affect the other measurements in intended or unintended ways, e.g. via formation of traversable or non-traversable wormholes \cite{Gao:2016bin,Maldacena:2017axo,Maldacena:2018lmt}.

    \item What kind of operator algebraic structures allow for such replications of information to happen in a consistent framework and including the simple observables of a smooth semi-classical geometry? Here we will need to study the details of the islands (entanglement wedges in the interior) and their semi-classical time-dependent behavior along with their interaction with the quantum hair and the Hawking radiation.
\end{itemize}

To explore the first question above, we will need to also quantize the bulk scalar fields representing the infalling matter in the interior explicitly. Crucial insights can come from checking the validity of the quantum null energy condition \cite{Bousso_2016} which has been shown to implement principles of quantum thermodynamics in holographic setups \cite{Kibe:2021qjy,Banerjee:2022dgv}. To explore the second direction, we will need to build on some recent developments which have explored the algebraic structures for semi-classical gravity and also quantum gravity in lower dimensions \cite{Leutheusser:2021frk,Witten:2021unn,Penington:2023dql,Kolchmeyer:2023gwa,AliAhmad:2023etg,Jensen:2023yxy}.

The fundamental question of the emergence of the semi-classical geometries via decoherence in quantum gravity and its consequences for resolving the AMPS critique of the complementarity principle has been explored in \cite{Bao:2017who}. It has been pointed out here that while the newly emitted Hawking quanta could be maximally entangled within an effective description constructed in a specific semi-classical geometry (or rather in a local semiclassical description in our model), the early and late Hawking radiation could be entangled in the global Hilbert space and not in a specific semi-classical branch of the wave-functional of the collapsing and evaporating black hole. In order to realize such a scenario, it is important to have some degrees of freedom within each branch of the wave-functional that can act as pointers as in the case of decoherence via interaction with the environment. Our model could provide an explicit way to test such a scenario and its consistency with the algebraic framework of quantum gravity. The hair degrees of freedom in our model that are approximately split into those which decouple from and those which get locked to the interior can act as the pointers since they satisfy the split property between interior and exterior of an evaporating black hole (like the semi-classical geometry) after scrambling time at least for simple observables.

Nevertheless, in order to fully investigate these issues, we should quantize the JT gravity system with matter in each throat following \cite{Penington:2023dql,Kolchmeyer:2023gwa} and study the full problem with quantum infalling matter. This could be feasible. It is likely that even in a given branch of the wave-functional, a semi-classical description can strictly apply locally and globally only if simple measurements are performed. Also such pointer bases of the wave-functional may emerge naturally like the energy eigenbases in the context of Eigenstate Thermalization Hypothesis \cite{DAlessio:2015qtq} so that, for typical initial conditions, the semi-classical observables are approximately diagonal in such a basis up to corrections that mimic a suitable random matrix ensemble and which are exponentially suppressed by the von-Neumann entropy. See \cite{Jafferis:2022wez} for such an explicit construction in the context of semi-classical gravity. The generalization of our model with each two-dimensional throat fully quantized can enable an explicit study of the emergence of semi-classical gravity \textit{along with the complementarity principle}, and lead to its fuller understanding from the algebraic point of view.

We want to emphasize that our model has a dual description as a lattice of coupled and entangled SYK-type quantum dots. Therefore, we also plan to study protocols for realistic simulations of our model in the future. In this context, we also want to point out that our model is related to some models of strange metals based on lattices of SYK-type quantum dots \cite{Sachdev:2023fim}, and particularly those which have local (momentum-independent) self-energy and universal spectral functions \cite{Doucot:2020fvy,Samanta:2022rkx} that can originate fundamentally from the decoupling principle discussed in this work \footnote{The universal spectral function gives an elegant way to derive linear-in-T resistivity and a refined picture of Planckian dissipation analytically \cite{Samanta:2022rkx}.}.

To conclude, we would like to end with a speculation about how we can generalize our model to capture microstate dynamics of a Schwarzchild or an astrophysical black hole instead of a near-extremal one. Recently, the study of spectral form factors given holographically by a bulk scalar field in a black brane geometry with the stretched horizon replaced by an infrared cutoff with random boundary conditions has shown to produce the linear ramp as expected from the full black hole geometry \cite{Das:2023ulz}. It has been speculated that such a simple modification can capture properties of a typical fuzzball microstate of a Schwarzchild black hole. In our model, such finite-temperature effects might be incorporated by introducing end-of-the-world (EOW) branes in each throat (see \cite{Penington:2019kki} as for instance) which provides an infrared cut-off and also by choosing the quantum state of these EOW branes randomly to mimic a typical microstate. In order to examine the feasibility of such an approach, it would be important to see if such a construction can account for the Bekenstein-Hawking entropy quantitatively following methods developed recently in \cite{Balasubramanian:2022gmo,Balasubramanian:2022lnw}.

\begin{acknowledgments}
It is a pleasure to thank Rajesh Gopakumar, Alok Laddha, Prabha Mandayam, Giuseppe Policastro, Suvrat Raju, Ashoke Sen and  Sai Vinjanampathy for many helpful discussions. The research of TK is supported by the Prime Minister's Research Fellowship (PMRF). AM acknowledges the support of the Ramanujan Fellowship of the Science and Engineering Board of the Department of Science and Technology of India, the new faculty seed grant of IIT Madras and the additional support from the Institute of Eminence scheme of IIT Madras funded by the Ministry of Education of India. HS acknowledges support from the INSPIRE PhD Fellowship of Department of Science and Technology of India.
\end{acknowledgments}

\appendix
\section{A black hole and a classical oscillator}
\label{Sec:AppendixA}
In this section we study the classical limit of the harmonic oscillator in the single throat model.

\begin{figure}
        \centering
        \includegraphics[scale=0.7]{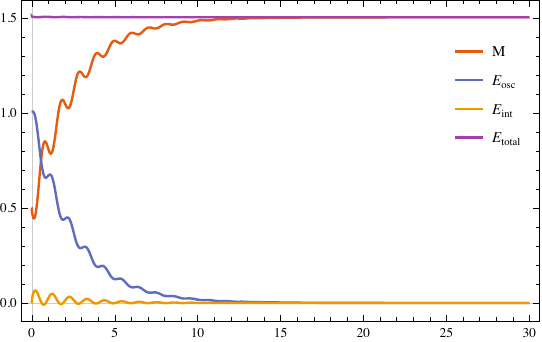}
        \caption{ $E_{\text{osc}} =  \frac{1}{2} \dot{x}^2 + \frac{1}{2} \omega^2 x^2$ and $E_{\text{int}} =  -(\Delta - 1) \gamma O x $. We have chosen $\omega =3.2 N^2$, $\gamma = -0.3 N$ and $\Delta=\frac{5}{4}$.}
        \label{Fig:energy_class_osc}
\end{figure}

\begin{figure}
\centering     
\subfigure[]{\includegraphics[scale=0.7]{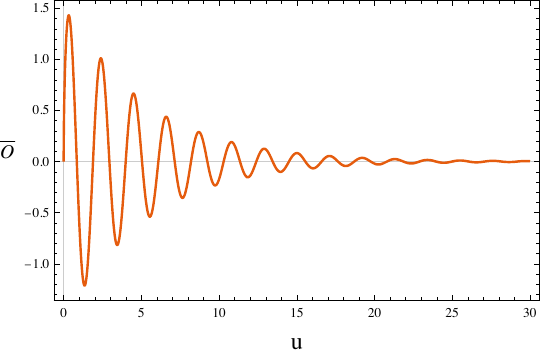}}
\subfigure[]{\includegraphics[scale=0.7]{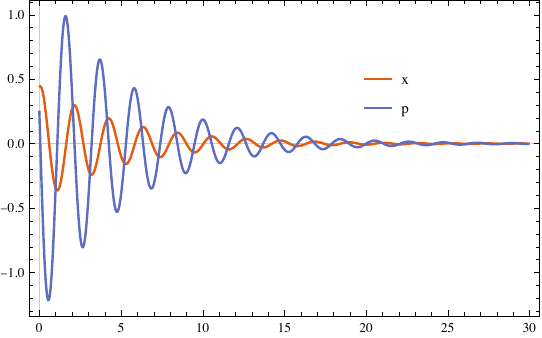}}
\caption{(a) $\bar{O}(u) =  \frac{O(u)}{N}$ decays to zero with time. (b) The position and momentum of the oscillator as functions of time.  We have chosen $\omega =3.2 N^2$, $\gamma = -0.3 N$ and $\Delta=\frac{5}{4}$.}
\label{Fig:osc_classical}
\end{figure}

The equations of motion are
\begin{align}\label{Eq:EOMClassOsc}
    \dot{M}(u) &= \frac{16\pi G}{\Phi_r}\gamma\left(\Delta O(u)  \dot{x}(u)+ (\Delta-1) \dot{O}(u) x(u)\right),\\
    \ddot{x} &= - \omega^2 x -\gamma O(u).
\end{align}
This model can be evolved numerically by generalizing the algorithm from \cite{Joshi:2019wgi} as described in Appendix \ref{Sec:AppendixB}. At initial time $u=0$ we choose the normalizable part of the bulk scalar as $\chi_n(r,0) = r^{-\Delta+1}\left(\chi(r,0) -J(0) r^{1-\Delta} -J'(0) r^{2-\Delta}\right) =0$ and $\partial_r\chi_n(r,0) =0$. This implies, from the near boundary expansion of the scalar, that $O(0) = 0$.
A typical evolution of energies is shown in Fig. \ref{Fig:energy_class_osc}.
As shown in Fig. \ref{Fig:energy_class_osc} the initial energy of the oscillator is completely eaten up by the black hole (blue and red curves) with the total energy conserved (magenta curve). The interaction energy (yellow curve) decays to zero at late times. We therefore see that the black hole and oscillator eventually decouple from each other. This is unsuprising in the classical case, since the black hole simply eats up all the energy of the oscillator. Remarkably this dynamical decoupling is a generic feature of such systems and will persist even when the oscillator is quantized as discussed in the main text. The amplitude of the position of the oscillator and the expectation value of the dual operator $O(u)$ decays to zero as shown in Fig.\ref{Fig:osc_classical}. 

\begin{figure}
\centering     
\subfigure[]{\includegraphics[scale=0.7]{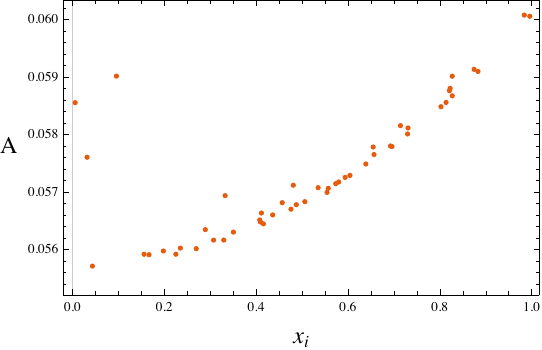}}
\subfigure[]{\includegraphics[scale=0.7]{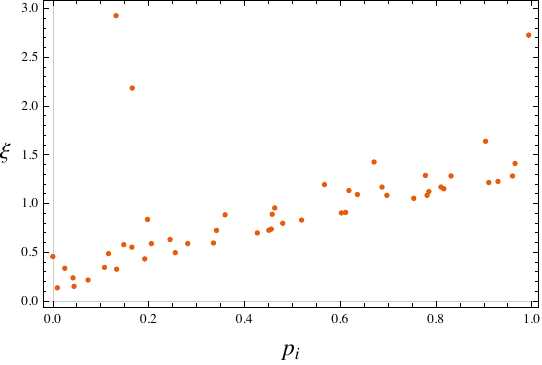}}
\caption{(a) The amplitude of the ringdown vs the initial position $x_i$ for different initial conditions. (b) The phase of the ringdown vs the initial momentum $p_i$ for different initial conditions.  We have chosen $\omega =3.2 N^2$, $\gamma = -0.3 N$ and $\Delta=\frac{5}{4}$. The outliers are largely due to the errors involved in computing numerical Fourier transforms of the data. }
\label{Fig:encode_class_osc}
\end{figure}

The oscillator behaves essentially like a damped harmonic oscillator and it forgets its initial state (position and momentum at $u=0$). This information is not completely lost but is encoded into the late time ringdown of the black hole mass.  as shown in Fig.\ref{Fig:encode_class_osc}.

\section{Algorithm for time evolution}
\label{Sec:AppendixB}
\subsection{Algorithm for the single throat model}
In this section we summarize the numerical algorithm used to evolve the black hole coupled to the oscillator, which builds on the algorithm in \cite{Joshi:2019wgi}. To illustrate the algorithm we consider the example of the classical oscillator coupled to the black hole with the following equations of motion
\begin{align}
    \dot{M}(u) &= \frac{16\pi G}{\Phi_r}\gamma\left(\Delta O(u)  \dot{x}(u)+ (\Delta-1) \dot{O}(u) x(u)\right),\\
    \ddot{x} &= - \omega^2 x -\gamma O(u).
\end{align}
The black hole mass and the position of the oscillator $x(u)$ need to be evolved self-consistently with the Klein-Gordon equation for the bulk scalar field, which is
\begin{equation}
    \partial_r\left( d_+ \chi \right) + \frac{\Delta (\Delta-1)}{2 r^2} \chi =0.
\end{equation}
where
\begin{equation}
    d_+ \chi(r,u) = \left(\partial_u  -\frac{1}{2}\left(1- r^2 M(u)\right) \partial_r \right)\chi(r,u).
\end{equation}
The near boundary expansion of the scalar field is
\begin{equation}
    \chi(r,u) \approx \chi^{(0)}(u) r^{1-\Delta} +\dot{\chi}^{(0)}(u) r^{2-\Delta} + O(u)r^{\Delta} + \hdots,
\end{equation}
It is useful to define
\begin{equation}
    \overline{d_+ \chi} = d_+ \chi - \frac{\Delta-1}{2} \chi^{(0)}(u) r^{-\Delta} -\frac{\Delta}{2} r^{1-\Delta} \dot{\chi^{(0)}}(u),
\end{equation}
which has a finite near boundary expansion
\begin{equation}
     \overline{d_+ \chi} \approx -\frac{\Delta}{2}r^{\Delta-1} O(u).
\end{equation}
Note that
\begin{equation}
    \partial_u \chi =  \overline{d_+ \chi} +\frac{\Delta-1}{2}  \chi^{(0)}(u) r^{-\Delta} +\frac{\Delta}{2} r^{1-\Delta} \dot{ \chi^{(0)}}(u)+\frac{1}{2}\left(1- r^2 M(u)\right) \partial_r\chi(r,u).
    \label{Eq:appendDu}
\end{equation}
The equation of motion for $\overline{d_+ \chi}$ is
\begin{equation}
    \partial_r \overline{d_+ \chi} + \frac{\Delta (\Delta-1)}{2 r^2} \left(\chi - r^{-\Delta+1}  \chi^{(0)}(u)- r^{-\Delta+2} \dot{ \chi^{(0)}}(u)\right),
\end{equation}
from which we obtain
\begin{equation}
    \overline{d_+ \chi} = -\int_0^r d\Tilde{r}\frac{\Delta (\Delta-1)}{2 \Tilde{r}^2} \left(\chi - \Tilde{r}^{1-\Delta}  \chi^{(0)}(u)- \Tilde{r}^{2-\Delta} \dot{ \chi^{(0)}}(u)\right).
\end{equation}
It is easy to see that the integral on the right hand side of the above equation is finite. Thus, given an initial profile for the scalar field at $u=0$ we can obtain $\overline{d_+ \chi}$ at $u=0$. From this we can readily obtain $\partial_u \chi$, given initial data for $J(u) = \frac{1}{C_\Delta} \chi^{(0)}(u)$, using \eqref{Eq:appendDu}, which can be used to evolve the scalar field forward in time using the finite difference method. The asymptotic expansion for the scalar field can then be used to self consistently obtain $O(u)$. Finally, $O(u)$ can be plugged into the equations of motion for $M(u)$ and $x(u)$ to evolve them forward in time using the finite difference method. Note that unlike \cite{Joshi:2019wgi}, the algorithm described here solves the evolution of the system without recourse to the $SL(2,R)$ charges or a diffeomorphism from the time dependent metric \eqref{Eq:metricEF} to a metric with fixed mass. This is a crucial modification that leads to better numerical stability of the algorithm.

\subsection{Algorithm for the full microstate model}
Recall that the equations of motion for the microstate model are
\begin{align}
\dot{M}_i &= - \la(\bQ_{i-1}+ \bQ_{i+1}-2 \bQ_i)\dt \dot{\bq}_i - \frac{c}{24 \pi} M_{i},\\
\ddot{\bq}_i &=\frac{1}{\si^2}( \bq_{i-1} + \bq_{i+1} - 2 \bq_i) +  \frac{1}{ \la^2}(\bQ_{i-1}+ \bQ_{i+1}-2 \bQ_i).
\label{Eq:EOMappend}
\end{align}
These equations explicitly depend on the $AdS_2$ $SL(2,R)$ charges $\bQ_i$ and we therefore have to use the algorithm from \cite{Joshi:2019wgi} instead of the modified algorithm described in the previous sub-section.

Note that the metric for a single throat \eqref{Eq:metricEF} with a time dependent mass $M_i(u)$
\begin{equation}
     {\rm d}s^2 = -2 \frac{l^2}{r^2} {\rm d}u {\rm d}r - \left(\frac{l^2}{r^2} - M_i(u) l^2 \right) {\rm d}u^2,
\end{equation}
can be converted to a fixed $M=1$ mass metric
\begin{equation}
     {\rm d}s^2 = -2 \frac{l^2}{\rho^2} {\rm d}v {\rm d}\rho - \left(\frac{l^2}{\rho^2} -  l^2 \right) {\rm d}v^2,
\end{equation}
via the following diffeomorphism
\begin{equation}
    v=v_i(u), \quad \rho = \frac{\dot{v_i}(u) r}{1-\frac{\ddot{v_i}(u)}{\dot{v_i}(u)}r},
\end{equation}
with
\begin{equation}
    -2 Sch(v_i(u),u) +\dot{v_i}(u)^2 = M_i(u).
\end{equation}
In these coordinates the $AdS_2$ $SL(2,R)$ charges are given by
\begin{align}
    \bQ^0_i &= \frac{\dddot{v_i}}{\dot{v_i}^2} - \frac{\ddot{v_i}^2}{\dot{v_i}^3} - \dot{v_i}\\
    \bQ^{\pm}_i&= \left(\frac{\dddot{v_i}}{\dot{v}_i^2} - \frac{\ddot{v_i}^2}{\dot{v_i}^3} \mp \frac{\ddot{v}_i}{\dot{v}_i} \right) e^{\pm v_i},
\end{align}
and satisfy
\begin{align}
    \dot{\bQ}^0_i &= -\frac{1}{2 \dot{v}_i} \dot{M_i}\\
    \dot{\bQ}^\pm_i &=-\frac{e^{\pm v_i}}{2 \dot{v}_i} \dot{M_i}.
    \label{Eq:Qevolve}
\end{align}

The $u$ derivatives of the time reparametrization $v_i(u)$ can be written in terms of the $SL(2,R)$ charges at time $u$ as
\begin{align}
    \dot{v}_i &= \frac{1}{2} \left(\bQ^-_i e^{v_i} +\bQ^+_i e^{-v_i} - 2 \bQ^0_i  \right)\\
    \ddot{v}_i &=\frac{1}{4}\left( \bQ^-_i e^{v_i}-\bQ^+_i e^{-v_i} \right)\left(\bQ^-_i e^{v_i} +\bQ^+_i e^{-v_i} - 2 \bQ^0_i  \right)\\
    \dddot{v}_i &= \frac{1}{4}\left( \bQ^-_i e^{v_i}-\bQ^+_i e^{-v_i} \right) \left( {\bQ^-_i}^2 e^{2 v_i}+  {\bQ^+_i}^2 e^{-2 v_i} - \bQ^0_i\left( \bQ^-_i e^{v_i}-\bQ^+_i e^{-v_i} \right)  \right).
    \label{Eq:timesevolve}
\end{align}

We impose the initial conditions $v_i(u=0)= 0$ and the time derivatives of $v_i$ are specified in terms of the initial $SL(2,R)$ charges that describe a microstate. The equation of motion for $\dot{M}_i$ \eqref{Eq:EOMappend} at time $u$ can be combined with \eqref{Eq:Qevolve} to evolve the $SL(2,R)$ charges forward in time to $u+\delta u$ using the finite difference method. The $SL(2,R)$ charges at time $u$ are used to evolve the time reparametrization \eqref{Eq:timesevolve} forward in time to $u+\delta u$ using the finite difference method. Finally the hair charges can be readily evolved forward in time using the equation of motion \eqref{Eq:EOMappend}.

The main numerical limitation of this method is the appearance of $\dot{v}_i$ in the denominator of \eqref{Eq:Qevolve} since the time reparametrization $v_i(u)$ always evolves such that $\dot{v}_i$ becomes small at late times leading to the amplification of noise. This issue could be avoided for the single throat model described in the previous sub-section since the model doesn't make explicit recourse to the $SL(2,R)$ charges. However, this numerical issue cannot be easily circumvented for the full microstate model.

\section{Data fitting for the black hole ringdown}
\label{Sec:AppendixC}
In this section we describe the procedure used to extract features, such as the amplitude and phase, of the late time black hole ringdown. As described in the main text, the late time ringdown is well described by
\begin{equation}
    M(u) = M_f - e^{-\tau u} \left(B \sin(\Omega u + \xi) + C \right).
\end{equation}

\begin{figure}[h]
\centering
\includegraphics[scale=0.7]{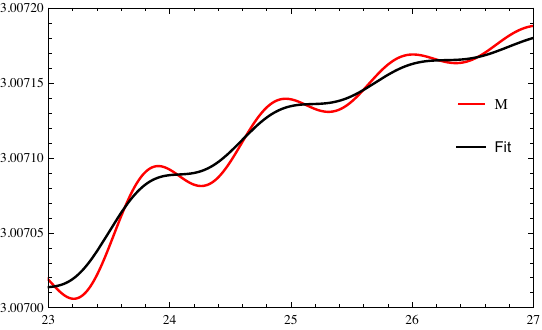}
\caption{A typical fit (black curve) of the late time mass (red curve) for a classical oscillator coupled to the black hole.}
\label{Fig:ringfit}
\end{figure}

To fit the mass data to this function we first extract the exponential envelope as follows. We apply a max filter followed by a Gaussian filter to the late time data for $M_f - M(u)$. This filters out the oscillating part of the ringdown, allowing us to extract $\tau$ by fitting the filtered data to a decaying exponential function. The oscillating part of $M_f - M(u)$ can be isolated by considering 
\begin{equation}
    \exp(\log(M_f-M(u)) +\tau u) = B \sin(\Omega u + \xi) + C.
\end{equation}
We then perform a discrete Fourier transform of the data for $\exp(\log(M_f-M(u)) +\tau u)$, which yields $B$, $C$, $\Omega$ and $\xi$. A typical fit of the late time mass ringdown is shown in Fig.\ref{Fig:ringfit}

\section{Open boundary conditions for $AdS_2$}
\label{Sec:AppendixD}
In this section we briefly review how Hawking radiation can be studied in an $AdS_2$ black hole by coupling the black hole to a non-gravitating bath region and imposing open boundary conditions such that the radiation is collected into the bath, following \cite{AEMM}.

\begin{figure}[h]
\centering
\includegraphics[scale=0.7]{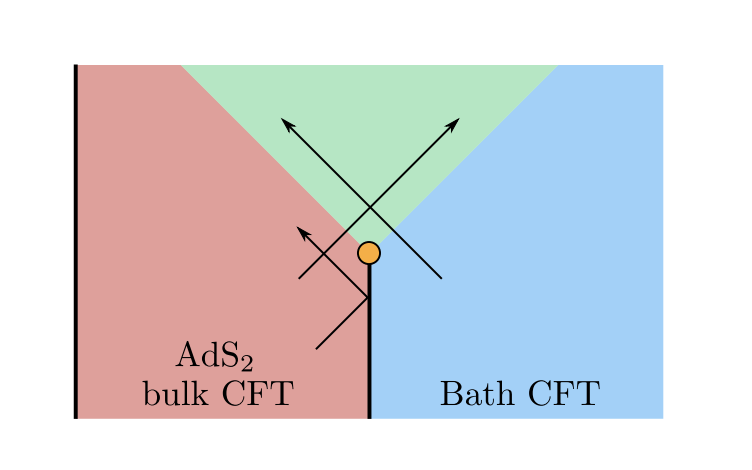}
\caption{Schematic of the setup to incorporate Hawking radition. This figure is from \cite{AEMM}.}
\label{Fig:AEMM}
\end{figure}

Initially, for $u<0$, $AdS_2$ has reflecting boundary conditions indicated by solid black lines in Fig.\ref{Fig:AEMM}. We consider conformal matter in the bulk described by a CFT with central charge $c$. A copy of the same CFT describes the bath. At $u=0$ (orange dot in Fig.\ref{Fig:AEMM}) we impose transparent boundary conditions, coupling the bulk CFT to the bath CFT. Note that we assume that the bulk CFT doesn't couple to the hair in the full microstate model.

As described in the main text \eqref{Eq:EOMJT1}, the bulk metric is locally $AdS_2$ and we can adopt Poincar\'e coordinates with the metric
\begin{align}
    ds^2 &= -\frac{4dx^+ dx^-}{(x^+ - x^-)^2} = \frac{4dx d\bar{x}}{(x + \bar{x})^2}\\
    \text{with, } &x^+ = t+ z = \bar{x}, x^- = t-z = -x
\end{align}
At $t=0$ the bulk matter in $AdS$ is assumed to be in the Hartle-Hawking state which can be conformally mapped to the vacuum on the half line $z>0$ with conformally invariant boundary conditions at $z=0$ \cite{AEMM}. The bath is also assumed to be in the same state at $t=0$ when the open boundary conditions are imposed. Note that the physical time $u$ is not the same as the Poincar\'e time $t$. As described in the main text the Poincar\'e time is related to $u$ via a time reparametrization as $t = t(u)$, where we can choose $t(0) =0$. The bath at time $u$ must be coupled to the boundary of $AdS$ at the same physical time corresponding to the Poincar\'e time $t(u)$.

As described in \cite{AEMM} we can use a diffeomorphism to map the initial state of matter on the $t=0$ slice, that includes both $AdS$ and the bath, to the vacuum state on the half line parametrized by $w \in [0,\infty)$. The expectation value of the stress tensor vanishes in the half-line vacuum implying that $\braket{T_{ww}(w)} = 0$. In $AdS_2$ the stress tensor expectation value vanishes in the Poincare metric and also in the Weyl rescaled flat metric $dx d\bar{x}$ since the Weyl anomaly between $AdS_2$ and flat space vanishes in Poincare coordinates. Thus, $\braket{T_{xx}(x)} = 0$. Now using the transformation of the stress tensor we obtain \cite{AEMM}
\begin{equation}
   \left(\frac{dw}{dx}\right)^2 \braket{T_{ww}} =  \braket{T_{xx}} +\frac{c}{24 \pi} Sch(w(x),x).
   \label{Eq:stressanomaly}
\end{equation}
Since $\braket{T_{ww}} = \braket{T_{xx}} = 0 $, the diffeomorphism $w(x)$ must be a $SL(2,R)$ transformation. This is chosen such that the $AdS_2$ region is mapped to $w\in[0,w_0)$ and the bath region is mapped to $w\in (w_0,\infty)$. The diffepmorphism is the following
\begin{equation}
	w(x) = \begin{cases}
			\frac{w_{0}^{2} \dot{t}(0)}{x + \dot{t}(0) w_{0}},& x >0\\
			w_{0} + t^{-1}(-x), & x<0
		\end{cases}
\end{equation}
This is a slight generalization of the diffeomorphim considered in \cite{AEMM}, ensuring that $w(x)$ and its first derivative are continuous at $x=0$. The discontinuity in the second derivative of $w$ leads to the initial delta function shock, mentioned in the main text, as evident from \eqref{Eq:stressanomaly}. This setup is repeated for each throat in the microstate model. Note that the value of $\dot{t}(0)$ for each throat will be determined by the initial values of the $SL(2,R)$ charges (see \eqref{Eq:Qevolve}). One can readily see that anomaly in \eqref{Eq:stressanomaly} leads to the modification of the equations of motion for the microstate model \eqref{Eq:LatticeEoms4}.

\bibliographystyle{apsrev4-1}
\bibliography{NAdS2LN}

\end{document}